\documentclass{aa}  
\usepackage{graphicx}
\usepackage{txfonts}
\usepackage{hyperref}
\bibpunct{(}{)}{;}{a}{}{,}

\usepackage{caption}
\usepackage{subcaption}
\usepackage[dvipsnames]{xcolor}
\usepackage{tabu}
\usepackage{placeins}

\begin{document}

        \title{Polycyclic aromatic hydrocarbon (PAH) abundances in the disk around T Chamaeleontis}
        \subtitle{PAH sizes, ionisation fraction, and mass from JWST observations}
        
        \author{Rahul Bandyopadhyay
                \inst{1,2}
                \and
                Simon Casassus\inst{1,3}
        }
        
        \institute{Departamento de Astronom\'{i}a, Universidad de Chile, Casilla 36-D, Santiago, Chile\\
                \email{rbandyo@das.uchile.cl}
                \and
                Millennium Nucleus on Young Exoplanets and their Moons (YEMS), Chile
                \and
                Data Observatory Foundation, Eliodoro Y\'{a}\~{n}ez 2990, Providencia, Santiago, Chile
        }

    \titlerunning{PAH abundances in T Cha}
    \authorrunning{Bandyopadhyay and Casassus}

        \date{Received Month 00, 0000; accepted Month 00, 0000}
        
        \abstract 
        {The T Tauri star T Chamaeleontis (T Cha) is known to have a protoplanetary disk (PPD) with a dust gap separating the inner and outer disk regions. The mid-infrared (MIR) JWST spectrum of T Cha shows multiple prominent aromatic IR bands (AIBs) around 6.2, 8.1, and 11.3 $\mu$m. AIBs are commonly accepted as the emission stemming from polycyclic aromatic hydrocarbon (PAH) molecules. Although PAHs may be instrumental for disk processes involving gas heating, thermochemistry, and photoevaporation, their abundances in PPDs are not well understood thus far.}         
        {We aim to characterise the PAHs giving rise to the AIBs observed in the JWST spectrum of T Cha. Our objective is to estimate the PAH abundances, in terms of their sizes, ionisation fraction, and mass, in the PPD of T Cha.}
        {We primarily used the archival MIRI-MRS JWST spectrum of T Cha for this work. We performed a spectral fitting of the observed AIBs to identify the possible underlying PAH emission components. We transferred the stellar radiation through a parametric disk model to reproduce the observed MIR continuum and the AIB features, as well as the optical photometric and millimetre band fluxes of T Cha. We included PAH dust grains, which were stochastically heated in our model calculations to simulate the AIBs. Thus, we were able to estimate the PAH abundances from our modelling. We used the results from previous observations and modelling efforts to reduce our model degeneracies.}
        {We estimated the PAH abundances in T Cha, along with other important disk parameters, from our modelling. The overall disk morphology derived in this work, made up of an inner and an outer disk separated by a dust gap, is consistent with the previous results from Spitzer, VLT, and ALMA observations. PAHs are located within the outer disk in our model. Given our best fiducial model, we estimated a population of small PAHs of $\lesssim26$ C atoms, with an ionised PAH fraction of $\sim0.15$. We also obtained a PAH-to-dust mass ratio of $\sim7\times10^{-3}$, which amounts to $\sim17\%$ of the PAH-to-dust mass ratio observed in the ISM. We predicted that the outer disk would  have a frontal wall with smaller dust grains (of sizes limited up to $\mu$m-order) to properly fit the continuum slope within $14-25$ $\mu$m. In our model, we propose the possibility of sub-micron dust grains within the gap, to justify an observed plateau around $\sim$10 $\mu$m in the JWST spectrum.}   
        {While our models do exhibit degeneracies, we have been able to predict a population of smaller and mostly neutral PAHs in the disk of T Cha. We also suggest that the disk might be undergoing FUV photoevaporation based on the high PAH-to-dust mass ratio estimated from our modelling.}   
            
        \keywords{circumstellar matter -- protoplanetary discs -- stars: individual: T Cha -- radiative transfer}
        
        \maketitle
        %
        
        \section{Introduction} \label{sec:introduction} 
        
        The infrared (IR) spectra of most astrophysical objects reveal a group of emission features, commonly around 3.3, 6.2, 7.7, 8.6, and 11.3 $\mu$m. It has been long established that a family of very small (predominantly) aromatic carbonaceous species are responsible for the emission \citep{LegerPuget1984, Allamandola+1989} and, hence, these features are generally referred to as aromatic IR bands (AIBs, e.g. \citealt{Verstraete+2001}). The small species have a lower heat capacity, so they are transiently heated and undergo steep temperature excursions following the absorption of single ultraviolet (UV) photon \citep{Sellgren1984}. Such a stochastic nature of heating makes it possible to reach temperatures that are about one order higher than in cases of equilibrium heating, even far from the radiation source; thus, they might significantly influence processes within any astrophysical medium.  
        
        Polycyclic aromatic hydrocarbon (PAH) molecules are widely considered to be the principal carriers of AIBs, noting their C-C and C-H vibrational modes coincide with the observed AIB wavelengths 
        (e.g. \citealt{Tielens2008, JoblinTielens2011, Li2020, Peeters+2021, Hrodmarsson+2025}). For example, the 3.3 $\mu$m band correspond to the C-H stretching within a PAH molecule, while the 6.2 and 7.7 $\mu$m bands correspond to the C-C stretching. The 8.6 $\mu$m band arise due to C-H in-plane bending, while the 11.3 and 12.7 $\mu$m bands generate due to C-H out-of-plane bending (e.g. \citealt{Peeters+2002}). However, it has not  been possible thus far to identify individual PAH molecules responsible for an AIB (e.g. \citealt{Li2020}). PAHs are instrumental in the physics and chemistry of their environment through the photoprocessing of the far-UV (FUV) photons \citep{BakesTielens1994, HollenbachTielens1997}. 
        
        In protoplanetary disks (PPDs), PAHs are dynamically coupled to the gas and trace the disk surfaces (e.g. \citealt{Lagage+2006}). Emission from the transiently heated PAHs are usually extended up to the outer disks (e.g. \citealt{Habart+2006, Yoffe+2023}). Thus, PAHs could play a critical role in gas heating and overall disk thermodynamics, including chemistry propagation, through the FUV irradiated disk surfaces, even in the outermost disk regions (e.g. \citealt{KampDullemond2004, Visser+2007, Kokoulina+2021}). PAHs could further influence the photoevaporative evolution of the disk through kinetic heating of gas, as models predict that the FUV photoevaporation efficiency increases in the presence of PAHs \citep{GortiHollenbach2009, Gorti+2009}. Therefore, knowledge of the PAH abundances in PPDs is important for a complete understanding of the disk processes during planet formation stages. 
        
        Based on Spitzer observations, one or more AIB features were detected in $\sim$70$\%$ of the PPDs around Herbig Ae/Be stars \citep{Acke+2004, Acke+2010}. However, only $\sim$10$\%$ of the disks around T Tauri stars have shown such features \citep{Geers+2006, Geers+2009}, which might be due to the overall UV faintness of the T Tauri stars. However, the major non-detections in T Tauri disks as well as those in Herbig Ae/Be disks might also be due to PAH cluster formation and subsequent adsorption into larger grain surfaces, which could  suppress the PAH emission and, hence, the AIBs \citep{Bakes+2001, Lange+2023}. Furthermore, PAHs could also be destroyed by EUV or X-rays \citep{SiebenmorgenKrugel2010}. {\cite{Geers+2006} used a standard radiative transfer model to fit the 11.3 $\mu$m band flux of a sample of T Tauri disks. They found a better fit to the overall sample considering around 10 to 100 times depletion of the PAH-to-dust mass ratio in their model compared to that of the ISM. The question of whether the PAH emissions were not detected or the PAHs were depleted due to processes related to disk evolution is not well-understood thus far. This makes the PAH abundances among PPDs difficult to constrain (e.g. \citealt{SeokLi2017}). 
        
        In this paper, we study the PPD around the T Tauri star T Chamaeleontis (T Cha), which have shown very prominent AIB features in its recent JWST observations. We interpret the AIBs as PAH emission profiles and we aim to model them to estimate the abundances of PAHs in the disk of T Cha. In general, the interstellar AIBs have been frequently interpreted in terms of dust emission from a population of very small PAH grains included within the interstellar dust models (e.g. \citealt{Desert+1990, LiDraine2001, Compiegne+2011}). These models thus provide optical properties of the PAH grains that can be fed into dust radiative transfer calculations to model the observed AIBs.
    
    We note a few studies that provide estimates of the PAH properties in PPDs through 3D dust radiative transfer modelling. These studies attempt to fit the spectral energy distributions (SEDs) and AIB features simultaneously in individual PPDs. For example, \cite{Woitke+2019} constrained the PAH mass and ionization fraction in 14 disks assuming single-sized population of PAH grains, which were considered to be at an equilibrium temperature different than the rest of the dust population. Their modelling included satisfactory fits to the prominent AIBs in HD 97048 and HD 169142. They found PAH-to-dust mass ratios in a range of $\sim0.1-25\%${\footnote{The percentage depends on the reference value of the ISM PAH-to-dust mass ratio. To avoid ambiguity, whenever discussing the fraction of PAH-to-dust mass ratio compared to that of the ISM, we considered an ISM PAH-to-dust mass ratio of $\sim4\times10^{-2}$, based on \cite{DraineLi2007}}} of the ISM PAH-to-dust mass ratio among their modelled disks, along with PAH ionization fractions of $\sim0.6-0.98$. 
    
    \cite{Sturm+2024} followed a simple treatment of PAHs in radiative equilibrium to match the observed strength of the AIB features in the edge-on disk HH48 NE. They considered fixed size distribution of the PAHs and obtained a PAH mass consistent with that of the ISM. The models from \cite{Maaskant+2014} are focussed on studying the PAH ionization across the disk gaps using a single-sized PAH population. They predicted that the ionization fraction might be larger in the optically thin gas within the gaps by modelling HD 97048, HD 169142, HD 135344 B, and Oph IRS 48.
    
    \cite{Habart+2021} modelled the AIBs observed in HD 100546 in terms of aromatised hydrocarbon nano-grains that show emission similar to PAH grains \citep{Jones+2013}. The nano-grains are stochastically heated in their model and have a fixed size distribution with a mass equivalent to $\sim$10$\%$ of the ISM PAHs. \cite{Devinat+2022} performed a similar treatment for HD 169142. 
     
    Focussed studies aimed at estimating PAH sizes, ionisation fractions, and masses simultaneously through models of individual disks are still rarely reported in the literature. Most studies so far have used generalised assumptions when interpreting the PAH abundances. In particular, we note that the effects of varying PAH size distribution in individual models were not explored well. Here, we aim to estimate detailed PAH abundances, defined in terms of PAH sizes, ionisation fraction, and mass within the disk of T Cha. We plan to obtain these self-consistently by reproducing the MIR dust continuum and the AIBs observed in the JWST spectrum T Cha through tailored thermal dust radiative transfer modelling. 
                                
        In Sect. \ref{sec:tcha}, we briefly describe key observations of T Cha found in the literature. In Sect. \ref{sec:observations}, we provide details of the observational data used for this work. In Sect. \ref{sec:modelapproach}, we describe the general modelling procedure with basic model assumptions. We discuss our best model, with the optimization process and  final results in Sect. \ref{sec:resultsdiscussion}. Finally, we summarise our work and draw our main conclusions in Sect. \ref{sec:summaryconclusions}. 
        
        \begin{figure}
                \centering
                \includegraphics[width=\columnwidth]{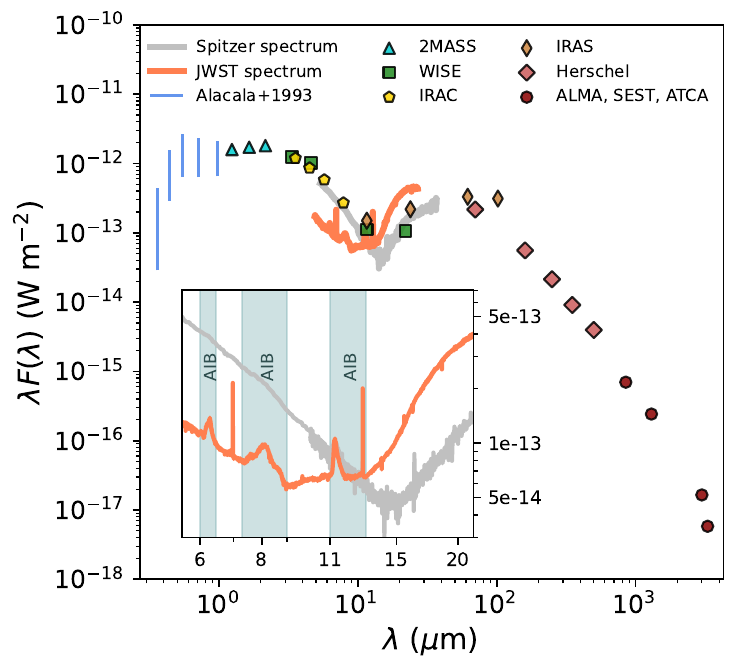}
                \caption{Collection of photometric fluxes and spectra of T Cha are shown. The blue vertical lines represent the variability in optical fluxes \citep{Alcala+1993}. IR photometric fluxes come from 2MASS, WISE, IRAC, IRAS, and Herschel. The millimetre band fluxes correspond to ALMA at 0.85 and 3 mm, SEST at 1.3 mm, and ATCA at 3.3 mm. The mid-IR spectra from Spitzer (grey) and JWST (orange) shows a varying continuum. Inset figure:  AIBs at 6.2, 8.1, and 11.3 $\mu$m, prominently appearing in the JWST spectrum, are marked. The two strong lines in the JWST spectrum correspond to the fine-structure [Ar~{\sc ii}] 6.98 $\mu$m and [Ne~{\sc ii}] 12.81 $\mu$m transitions.}
                \label{fig:obsdata}
        \end{figure}
        
        \begin{figure}
                \centering
                \includegraphics[width=7.9cm]{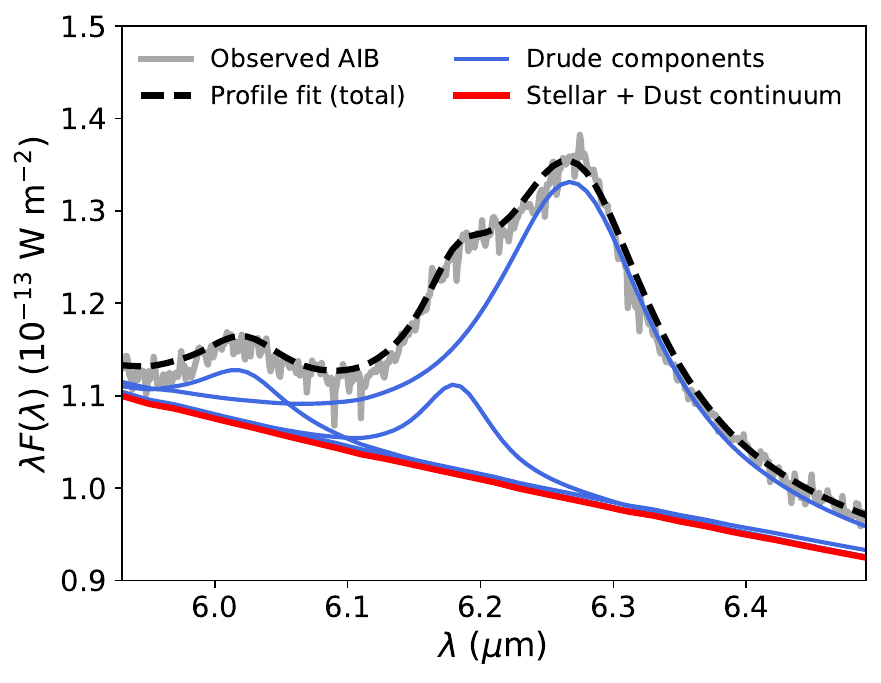}
                \includegraphics[width=7.9cm]{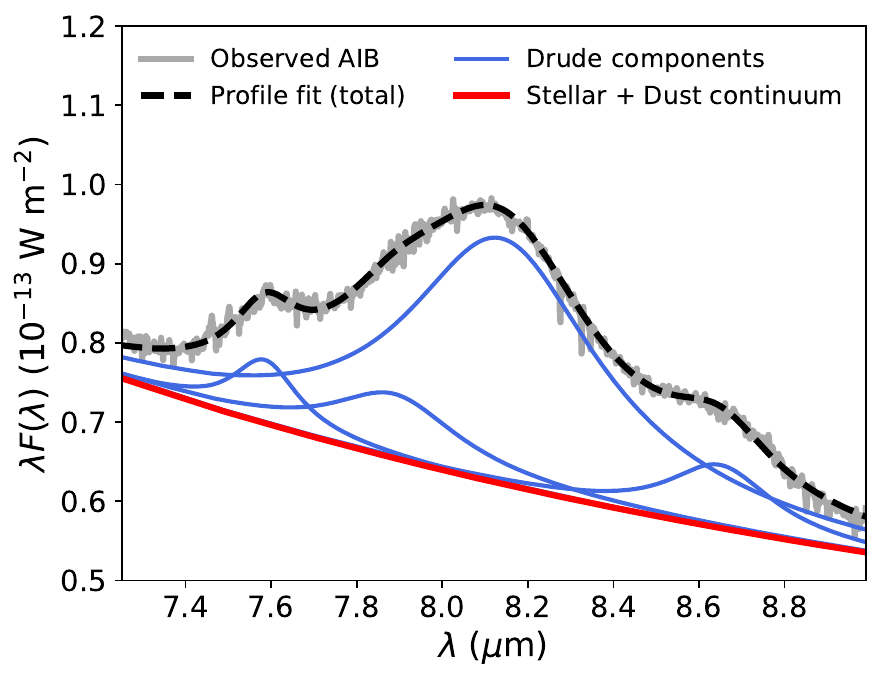}
                \includegraphics[width=7.9cm]{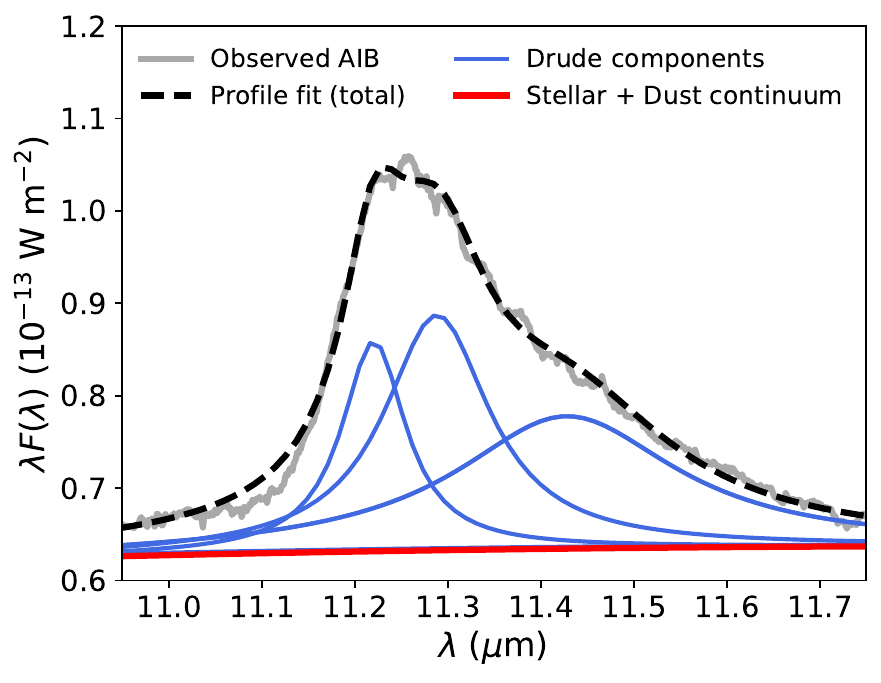}
                \caption{Spectral decomposition of the 6.2 (top), 8.1 (middle), and 11.3 (bottom) $\mu$m AIBs observed in the JWST spectrum of T Cha. Band profiles are fitted with Drude profile components (blue solid lines). The sum of the underlying stellar and dust continuum is shown in red solid line. The black dashed line shows the total fit to the AIB profiles. See more details in Sect. \ref{sec:pahprofiles} and Appendix \ref{app:nautilus}.}
                \label{fig:pahfit}
        \end{figure}

        \section{T Cha} \label{sec:tcha}
        
        T Cha has a spectral type of G8V \citep{Alcala+1993, Schisano+2009}, located at a distance of 103 pc (Gaia DR3; \citealt{GaiaDR3+2023}) within the $\epsilon$-Cha complex, and have an age of $\sim$7 Myr \citep{Torres+2008}. T Cha is one of the rare T Tauri disks, in which at least one AIB was detected from Spitzer observations \citep{Geers+2006}. The Spitzer spectrum depicted a dip around $\sim$10 $\mu$m, which was indicative of a dust gap. \cite{Brown+2007} reproduced the Spitzer spectrum by their radiative transfer model assuming a dust gap separating the inner and outer disks. Further characterizations of the T Cha disk has been done via a number of analyses and modelling interpretations of various data sets: VLTI/AMBER interferometry (\citealt{Olofsson+2011,Olofsson+2013}, hereafter O13); Herschel imaging \citep{Cieza+2011}; VLT/SPHERE polarimetry (\citealt{Pohl+2017}, hereafter P17); ALMA 0.8 mm (\citealt{Huelamo+2015}, hereafter H15) and 3 mm (\citealt{Hendler+2018}, hereafter H18) observations; and JWST \citep{Bajaj+2024, Sellek+2024, Xie+2025}. The presence of a possible planet within the cavity was predicted by \citet{Huelamo+2011} and later supported by P17 and H18.
        
        In earlier works, T Cha was shown to exhibit a variability of about three magnitudes in optical photometric fluxes \citep{Alcala+1993, Walter+2018}. The structure of the emission lines (e.g. H$\alpha$, [O~{\sc i}]) were also found to be variable from high-resolution spectroscopy \citep{Schisano+2009, Cahill+2019}. Recent JWST MIR spectrum of T Cha depicts a significant variation compared to its Spitzer spectra, as \cite{Xie+2025} reported up to a three-fold decrease in the continuum shortwards of $\sim$10 $\mu$m, with an equal amount of increase longwards of 10 $\mu$m. This kind of variability is known as the `see-saw' variability and was previously observed in multi-epoch Spitzer observations \citep{Kim+2009, Espaillat+2011}. This variability has been suggested to occur due to a reduction of the inner disk dust mass and/or scale height, reducing the inner disk SED contribution at shorter wavelengths. This simultaneously allows a greater amount of stellar radiation to reach the outer disk, increasing the outer disk SED in longer wavelengths. \cite{Xie+2025} modelled the disk SED to explain the variation between the Spitzer and JWST observations. They suggested an outburst event have reduced the inner disk dust mass and scale height in the case of T Cha. 
        
        \cite{Bajaj+2024} reported a spatially extended disk wind from T Cha, detected in ionic line emissions, from the JWST observations. The AIBs are also much prominently observed in the JWST spectrum compared to the Spitzer spectra (see more details on AIBs in Sect. \ref{sec:pahprofiles}). 
        
        \section{Observational data} \label{sec:observations}
        
        We retrieved the archival JWST data of T Cha published by \cite{Bajaj+2024data} in the VizieR On-line Data Catalog\footnote{\url{https://cdsarc.cds.unistra.fr/viz-bin/cat/J/AJ/167/127}}, originally available in the Mikulski Archive for Space Telescopes (MAST) (DOI: 10.17909/dhmh-fx64\footnote{\url{https://doi.org/10.17909/dhmh-fx64}}).  
        The spectra were taken on 13 Aug. 2022 as part of the programme 2260 (PI: Ilaria Pascucci) using the Mid-InfraRed Instrument (MIRI) onboard JWST in Medium Resolution Spectroscopy (MRS) mode. The observations were taken through four IFU channels operating in four different wavelength ranges covering the MIR region: $4.9-7.65$ $\mu$m (Channel 1), $7.51-11.7$ $\mu$m (Channel 2), $11.55-17.98$ $\mu$m (Channel 3), and $17.7-27.9$ $\mu$m (Channel 4). We also adopted the optical photometric fluxes reported in \cite{Alcala+1993}; ALMA band fluxes at 0.85 mm and 3 mm, presented by H15 and H18, respectively; and SEST 1.3 mm and ATCA 3.3 mm band fluxes from \cite{Lommen+2007}. 
        
        For comparison with the JWST spectrum (Fig. \ref{fig:obsdata}), we downloaded the Spitzer InfraRed Spectrograph (IRS) spectra of T Cha from the Combined Atlas of Sources with Spitzer IRS Spectra (CASSIS) \citep{Lebouteiller+2015}. The spectra were acquired in low-resolution mode: short-low (SH, $5.13-7.6$ $\mu$m) and high-resolution modes: short-high (SH, $9.9-19.6$ $\mu$m) and long-high (LH, $18.7-37.2$ $\mu$m) on 30 May 2005 (PI: James R. Houck; AORkey: 12679424). Additionally, we also collected photometric band fluxes from 2MASS, WISE, IRAC, IRAS, and Herschel catalogues available at the VIZIER Catalogue Service\footnote{\url{https://vizier.unistra.fr/}}.  
        
        \begin{table}
                \centering
                \caption{Decomposition of the AIB profiles.}
                \label{tab:pahflux}
                \begin{tabular}{lcccc}
                        \hline
                        AIB & $\lambda_\mathrm{j}$ & $b_\mathrm{j}$ & $\gamma_\mathrm{j}$ & FWHM$_j$\\
                        ($\mu$m) & ($\mu$m) & (Jy) & & ($\mu$m)\\
                        \hline
                        6.2 & 6.021 & 0.012 & 0.014 & 0.085\\ 
                        & 6.182 & 0.020 & 0.011 & 0.068\\ 
                        & 6.270 & 0.072 & 0.023 & 0.146\\ 
                        \hline
                        8.1 & 7.586 & 0.020 & 0.025 & 0.187\\ 
                        & 7.890 & 0.021 & 0.046 & 0.362\\ 
                        & 8.143 & 0.084 & 0.067 & 0.545\\ 
                        & 8.650 & 0.023 & 0.035 & 0.299\\ 
                        \hline
                        11.3 & 11.220 & 0.085 & 0.007 & 0.084\\ 
                        & 11.288 & 0.096 & 0.012 & 0.137\\ 
                        & 11.428 & 0.055 & 0.025 & 0.288\\  
                        \hline
                \end{tabular}
        \tablefoot{The bands are composed of multiple Drude profiles with profile parameters: central wavelength ($\lambda_\mathrm{j}$), fractional width ($\gamma_\mathrm{j}$), $\mathrm{FWHM}_\mathrm{j}(=\lambda_\mathrm{j}\gamma_\mathrm{j})$, and amplitude ($b_\mathrm{j}$).}
        \end{table}
        
        \section{Analyses of the {AIB} profiles} \label{sec:pahprofiles}
        
        A group of prominent AIBs could be seen on top of the MIR thermal continuum in the JWST spectrum of T Cha. The 11.3 $\mu$m band was prominent in the Spitzer spectra \citep{Geers+2006}. \cite{Arun2025} recently reanalysed the Spitzer spectra and reported detection of AIBs at 6.2, 7.7, and 8.6 $\mu$m. From the JWST spectrum, we can clearly identify the bands around 6.2 and 11.3 $\mu$m, while the 7.7 and 8.6 $\mu$m bands seem to have formed a complex centred around $\sim$8.1 $\mu$m. The 6.2 $\mu$m band is preceded by a weaker 6.0 $\mu$m band{\footnote{The 6.0 $\mu$m band is generally associated with a carbonyl ($>\mathrm{C} = \mathrm{O}$) stretch of a PAH molecule \citep{Peeters+2002}}}.
        
        \subsection{PAH spectral type}
         
        The AIB features vary in different astrophysical objects and environments, which led to their classification as $\mathcal{A}$, $\mathcal{B}$, and $\mathcal{C}$ by \cite{Peeters+2002}{\footnote{\cite{Peeters+2002} used the term `PAH' for classification}}; subsequently,  \cite{Matsuura+2014} added another class, $\mathcal{D}$. The classes are mainly based on the shapes and peak-positions of the 6.2, 7.7, and 11.3 $\mu$m profiles as well as the relative location of the 7.7 and 8.6 $\mu$m bands in terms of their central wavelengths. The class $\mathcal{A}$ and class $\mathcal{B}$ profiles show a clear separation between the 7.7 and 8.6 $\mu$m band profiles, while the peak of the 7.7 $\mu$m class $\mathcal{B}$ profile is $\sim$0.2 $\mu$m red-wards compared to that in class $\mathcal{A}$. Classes $\mathcal{C}$ and $\mathcal{D}$ exhibit a single profile in the 7-9 $\mu$m range, with the peaks located around 7.7 and 8.22 $\mu$m, respectively. Additionally, the class $\mathcal{D}$ category is suggested to have an aliphatic feature around 6.9 $\mu$m \citep{Matsuura+2014}.
        
        The JWST spectrum of T Cha depicts a single AIB profile within 7-9 $\mu$m peaking beyond 8 $\mu$m, which is predominantly identified with an overall spectral class $\mathcal{C}$. The exact peak location at 8.1 $\mu$m is slightly bluewards compared to the typical class $\mathcal{C}$ profile peak. 
        The profile also indicates a partial separation of the 7.7 and 8.6 $\mu$m bands by showing a red wing at 8.6 $\mu$m. This, along with a blue wing at 7.6 $\mu$m tentatively indicate towards a class $\mathcal{A}$/$\mathcal{B}$ contribution. The peak position of the 6.2 $\mu$m band, at 6.27 $\mu$m (as well as the profile shape) is further related to class $\mathcal{B}$ profiles. We noticed a very faint 6.9 $\mu$m feature, which is sometimes also present in classes $\mathcal{A}$ and $\mathcal{B}$ as well; hence, we could not draw any definitive conclusion for a class $\mathcal{D}$ contribution. Therefore, we find that T Cha may be qualitatively characterised as of PAH spectral class $\mathcal{C}$ with some contribution from class $\mathcal{B}$ signatures.
        
        An association of the class $\mathcal{A}$ profiles with highly UV processed environments, such as galaxies and ISM, can generally be observed, as compared to the prevalence of class $\mathcal{B}$, $\mathcal{C}$, and $\mathcal{D}$ profiles in circumstellar materials with lower UV exposure (e.g. \citealt{Li2020, Peeters+2021}). It is known that the aliphatic components of the hydrocarbon mixes are more susceptible to dissociation under higher UV irradiation than their more stable aromatic counterpart. Hence, objects depicting class $\mathcal{C}$ profiles, such as T Cha, could have a higher amount of aliphatic-to-aromatic ratio, which might be present in the form of a larger number of aliphatic sub-groups attached to the PAH molecules (e.g. \citealt{Sloan+2007, Dartois+2020}; see further discussion in Sect. \ref{sec:pahclassimplications} below).    
        
        \subsection{Profile fitting of the AIBs: spectral components}
        
        Despite the common understanding of PAH molecules giving rise to the AIBs, the individual molecules responsible for a particular AIB feature have not been identified so far (e.g. \citealt{Li2020}).
        However, detailed spectral fitting of the AIBs have improved the understanding of the features as combination of vibrational modes of PAHs \citep{Boulanger+1998, Verstraete+2001, Smith+2007}. 
        Furthermore, such analyses also formed the observational basis of deriving empirical opacities for PAH grains, for example, the `astro-PAH' opacities that are widely used to model dust emission \citep{LiDraine2001, DraineLi2007}. 
        
        The spectral fitting has been so far focussed on class $\mathcal{A}$ profiles (e.g. \citealt{Smith+2007}). Due to the major differences within 7-9 $\mu$m, the class $\mathcal{A}$ profile components are not applicable for T Cha with strong class $\mathcal{C}$ profile signatures. Hence, we attempted a spectral decomposition of the PAH profiles observed in the JWST spectrum of T Cha, in order to predict the underlying components in terms of their central wavelengths and width. We later utilised this information to derive PAH opacities that could be specifically applied to model T Cha (see further discussion on PAH opacities in Sect. \ref{sec:pahopacities}).  
        
        To fit the AIBs in T Cha and find out the probable underlying profile components, we mainly followed the technique from \cite{Smith+2007}, including the assumption that the vibrational modes of the PAH molecules can be theoretically approximated as that of a harmonic oscillator. Hence, we assumed the individual AIB profiles to be a combination of one or more Drude profiles of the functional form 
        \begin{equation}
                I_{\mathrm{D},j}(\lambda) = \frac{b_j\gamma_j^2}{\left( \lambda/\lambda_j - \lambda_j/\lambda \right)^2 + \gamma_j^2}.\\ 
                \label{eqn:1} 
        \end{equation}
        Here, $\lambda_\mathrm{j}$ and $\gamma_\mathrm{j}$ are the central wavelength and the fractional FWHM, respectively, of the $j$th profile, where $\mathrm{FWHM}_\mathrm{j}= \lambda_\mathrm{j}\gamma_\mathrm{j}$; $b_\mathrm{j}$ is the amplitude of the $j$th profile{\footnote{The integrated flux of the $j$th Drude profile is given by ${(\pi}c/2)b_\mathrm{j}\gamma_\mathrm{j}/\lambda_\mathrm{j}$ \citep{Smith+2007}}}.
        
    We fit each AIB profile in absolute flux scale with a function,
        \begin{equation}
                I(\lambda) = C(\lambda) + I_{\mathrm{PAH}}(\lambda);\\\\I_{\mathrm{PAH}}(\lambda)=\sum_j I_{\mathrm{D},j}(\lambda),\\ 
                \label{eqn:2} 
        \end{equation}
  where  $C(\lambda)$ represents the underlying continuum under the AIB profile. We express $C(\lambda)$ as the sum of the stellar spectrum, $S_*(\lambda)$, and the thermal dust continuum, which we can assume to be the combination of modified black body spectra. Hence,
        \begin{equation}
                C(\lambda) = S_*(\lambda) + \sum_k f_k B_\lambda(T_k)(\lambda/\lambda_0)^{-2},\\ 
                \label{eqn:3}
        \end{equation}
        where $f_\mathrm{k}>0$, $B_\lambda(T)$ denotes the black body function and $\lambda_0$ is the normalization wavelength assumed as 9.7 $\mu$m. 
        
        We employed a Bayesian inference technique to estimate the most probable Drude functions fitting the {AIB} profiles. For this purpose, we used the {\tt NAUTILUS} code \citep{Lange+2023nautilus} that runs a nested sampling algorithm to reach the maximum likelihood of the Drude profile parameters. We evaluated the likelihood function, $L$, defined as 
        \begin{equation}
                \log{L} = - \frac{1}{2} \sum_n \left[ \frac{(I_{\mathrm{obs},n} - I_{\mathrm{mod},n})^2}{\sigma^2} - \log(2\pi\sigma^2) \right],\\ 
                \label{eqn:4} 
        \end{equation}
        where the term within the square brackets is summed over $n$ number of observed points on the spectrum. $I_{\mathrm{obs},n}$ and $I_{\mathrm{mod},n}$ denote the observed and the modelled flux densities at the $n$th observed point, respectively.   
        Using this procedure, we find the best-fit Drude profile parameters, $\lambda_\mathrm{j}$, $b_\mathrm{j}$, and $\gamma_\mathrm{j}$. More details of the fitting procedure is described in Appendix \ref{app:nautilus}.  
         
        The 6.2 $\mu$m AIB in T Cha is fitted with two components at 6.182 and 6.27 $\mu$m. This is similar to the suggestion by \cite{Peeters+2002} that a 6.2 $\mu$m band having class $\mathcal{B}$ profile can be presented as the combination of class $\mathcal{A}$ and class $\mathcal{C}$. We note that the analysis of 6.2 $\mu$m band includes a fit to the 6.0 $\mu$m band since it is partially blended with the 6.2 $\mu$m profile. The 8.1 $\mu$m band in T Cha is found to be composed of four components, centred at 7.586, 7.89, 8.143, and 8.65 $\mu$m, based on our analyses (see Sect. \ref{sec:pahclassimplications}). We use three components at 11.22, 11.288, and 11.428 $\mu$m to satisfactorily fit the profile structure of the 11.3 $\mu$m AIB. The fit to each profile is shown in Fig. \ref{fig:pahfit}. The components derived for each AIB from this analyses are summarised in Table \ref{tab:pahflux} with their defining parameters.
        
        \subsection{Implications on aliphatic hydrocarbon bonds in T Cha} \label{sec:pahclassimplications}
        
        The 7.7 $\mu$m band was previously explained through observations as the combination of the sub-features at $\sim$7.6 and $\sim$7.8 $\mu$m by \citep{Peeters+2002}, where the dominance of the bluer sub-feature produces a class $\mathcal{A}$ profile while class $\mathcal{B}$ results when the redder component outshines. Since an aliphatic sub-group shifts the purely aromatic vibrational modes to longer wavelengths, the observed shift of the 7.7 $\mu$m band peak from class $\mathcal{A}$ to B has been explained in terms of an aliphatic contribution to the pure C-C modes that generate the 7.7 $\mu$m PAH emission \citep{Sloan+2007, ShannonBoersma2019}. When the $\sim$7.8 $\mu$m sub-feature largely dominates the $\sim$7.6 $\mu$m sub-feature, the 7.7 $\mu$m band nearly merges with the 8.6 $\mu$m band and might generate a class $\mathcal{C}$ profile (e.g. \citealt{Sloan+2007}). 
        
        From our analyses, we propose that the components of the 8.1 $\mu$m profile at 7.586 and 7.89 $\mu$m  specifically represent the class $\mathcal{A}$/$\mathcal{B}$ signatures of the 7.7 $\mu$m band and might correspond to purely C-C modes. The 8.65 $\mu$m component account for the well-known 8.6 $\mu$m C-H in-plane bending mode. The component at 8.143 $\mu$m may be interpreted as the link between the 7.7 and 8.6 $\mu$m bands that shapes the profile as a single feature within 7-9 $\mu$m. We hypothesise that the 8.143 $\mu$m feature collectively represents the emission from the PAHs with the contribution from the aliphatic sub-groups.

        \begin{figure}
                \centering
                \includegraphics[width=\columnwidth]{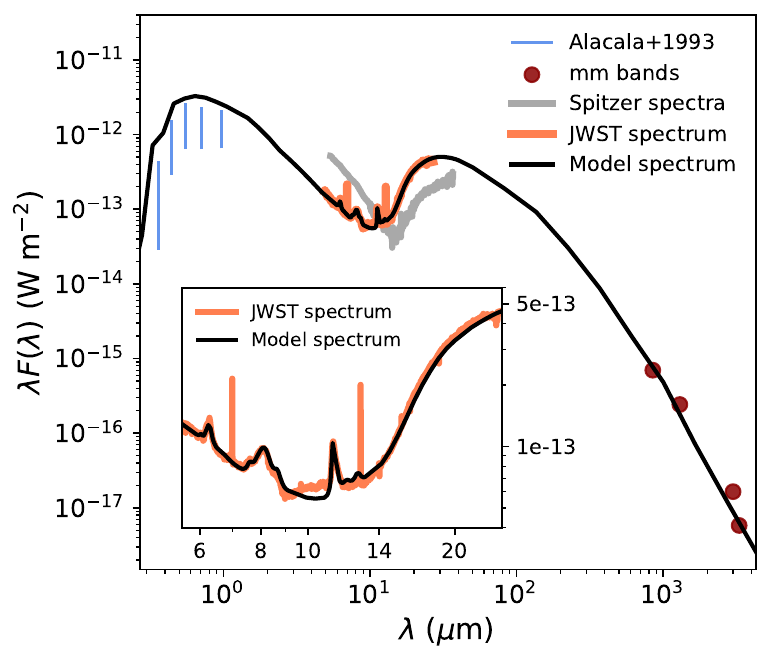}
                \caption{Model spectrum (black line, generated in a varying resolution across the wavelength range) is overplotted with the observed MIR JWST spectrum (orange line), optical photometric fluxes (blue vertical lines), and millimetre band fluxes (red circles). The MIR Spitzer spectrum (grey line) is shown as well for comparison. Inset plot: Fitting of the MIR continuum, along with the observed AIBs at 6.2, 8.1, and 11.3 $\mu$m, and the model generated continuum with the respective PAH bands, is shown in more detail.}
                \label{fig:model}
        \end{figure}
        
        \begin{table*}
                \centering
                \caption{Disk parameters from the best fiducial model}
                \label{tab:disk}
                \begin{tabular}{lccccc}
                        \hline
                        \multicolumn{6}{c}{Central star}\\
                        \hline
                        $T_\mathrm{eff}$ (K) &&&&& 5500\\
                        $L$ $(L_{\sun})$ &&&&& 1.39\\
                        $R$ $(R_{\sun})$ &&&&& 1.3\\
                        \hline
                        \hline
                        \multicolumn{6}{c}{Disk}\\
                        \hline
                        $i$ (deg) &&&&& 68\\
                        & \multicolumn{2}{c}{Inner disk} & Gap & \multicolumn{2}{c}{Outer disk}\\
                        & {Puffed-up} rim & Shadowed region & & Wall & Outer region\\
                        $R_\mathrm{in}$ (au) & 0.07 & 0.11 & 1 & 31 & 36\\
                        $R_\mathrm{out}$ (au) & 0.11 & 1 & 31 & 36 & 50\\
                        $\gamma$ & -1.0 & -1.0 & 0.0 & -1.0 & -1.0\\            
                        $H_0$ (au), $R_0$ (au) & 0.2, 1 & 0.05, 1 & 0.05, 1 & 1.9, 25 & 1.9, 25\\
                        $\beta$ & 1.0 & 1.0 & 1.1 & 1.1 & 1.1\\
                        $M_d$ ($M_{\sun}$) & $1.2\times10^{-12}$ & $1.9\times10^{-11}$ & $6\times10^{-10}$ & $7\times10^{-7}$ & $7.5\times10^{-5}$\\
                        \hline
                \end{tabular}
        \end{table*}
        
        \begin{table}
                \centering
                \caption{Model abundances of different dust species in separate disk regions.}
                \label{tab:dust}
                \begin{tabular}{lcccc}
                        \hline
                        & Species & $M_\mathrm{d}$ ($M_{\sun}$) & $a_\mathrm{min}$, $a_\mathrm{max}$  ($\mu$m) & $p$\\
                        \hline
                        Inner disk  & a-C  &  $3.0(-13)$   & 0.01, 1000 & -3.5\\
                        (puffed-up & a-Sil & $9.0(-13)$   & 5, 1000 & -3.5\\
                        rim) &&&&\\
                        \hline
                        Inner disk        & a-C  & $4.8(-12)$     & 0.01, 1000 & -3.5\\
                        (shadowed & a-Sil & $1.4(-11)$   & 5, 1000 & -3.5\\
                        region) &&&&\\
                        \hline
                        Gap                & a-C       & $1.5(-10)$  & 0.01, 1.0 & -3.5\\
                        & a-Sil     & $4.5(-10)$  & 0.01, 1.0 & -3.5\\
                        \hline                   
                        Outer disk     & n-PAH     & $4.3(-9)$  & 2.9(-4), 4.0(-4)    & -3.5\\
                        (wall)               & i-PAH     & $4.8(-10)$  & 2.9(-4), 4.0(-4)    & -3.5\\
                        & a-C       & $1.8(-7)$    & 0.01, 4.0 & -3.5\\
                        & a-Sil     & $5.3(-7)$    & 0.01, 4.0 & -3.5\\
                        \hline                   
                        Outer disk        & n-PAH     & $4.7(-7)$   & 2.9(-4), 4.0(-4)    & -3.5\\
                        (outer                 & i-PAH     & $5.2(-8)$   & 2.9(-4), 4.0(-4)    & -3.5\\
                        region) &&&&\\
                        & a-C       & $1.9(-5)$   & 0.01, 5000 & -3.3\\
                        & a-Sil     & $5.6(-5)$  & 0.01, 5000 & -3.3\\
                        \hline
                \end{tabular}
        \tablefoot{
                1. Dust abundaces are presented in terms of dust mass ($M_\mathrm{d}$); minimum ($a_\mathrm{min}$) and maximum ($a_\mathrm{max}$) radii of the dust grains; and the slope ($p$) of the power-law size distribution. The dust species include amorphous carbon: a-C, amorphous silicate: a-Sil, ionised PAH: i-PAH, and neutral PAH: n-PAH.\\
                2. The expression $X(Y)$ stands for $X\times10^Y$.
        }
        \end{table}

        \begin{figure}
                \centering
                \includegraphics[width=9cm]{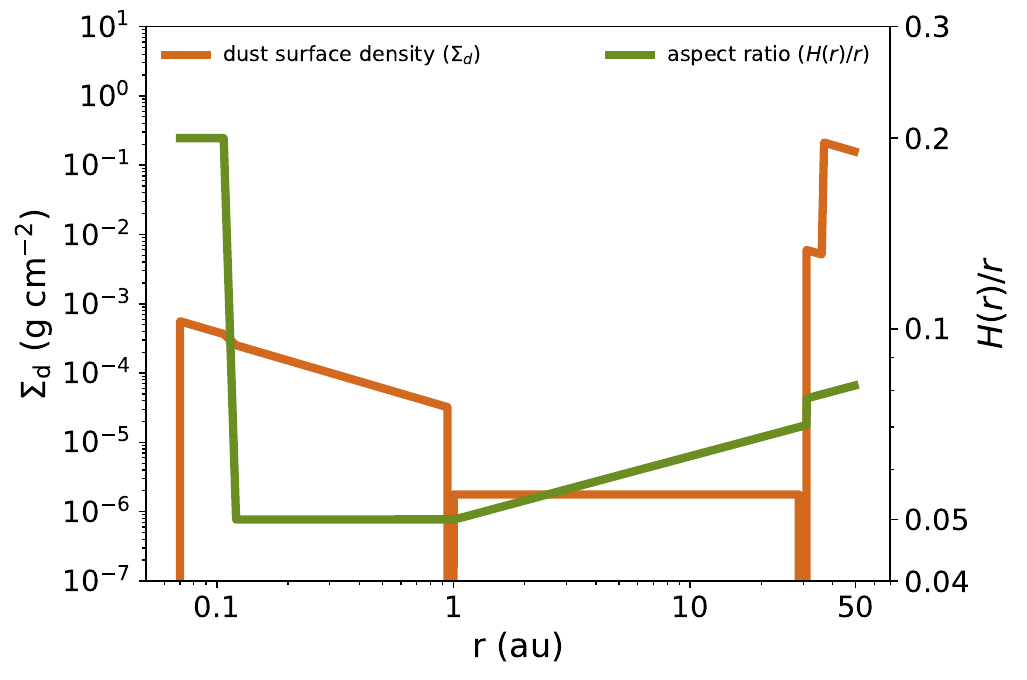}
                \caption{Radial profiles of the dust surface density ($\Sigma_\mathrm{d}$) and the aspect ratio (scale height to radius ratio, $H(r)/r$) of the modelled disk is shown in orange and green solid lines, respectively.}
                \label{fig:densityprofile}
        \end{figure}
        
    \section{Radiative transfer modelling} \label{sec:modelapproach}
        
        To estimate the PAH abundances in the disk of T Cha in a self-consistent way, we performed a dust radiative transfer modelling of the observed data using the code {\tt MCFOST} \citep{Pinte+2006}. The code considers propagation of a large number ($\sim10^5$) of photon packets, originating from a radiation source, through an input 3D density grid, adopting a Monte Carlo algorithm. It solves the radiative transfer problem at each grid cell and generates a temperature grid, which is further processed by the code to produce synthetic data, including images and/or an SED that can be compared to the observational data. We modelled only the dust emission from the disk of T Cha and did not focus on the line emission from gas. This section discusses the general modelling procedure with a few basic model assumptions applicable to T Cha.
        
        \subsection{Modelling aim} \label{sec:modelaim}
        
        We aimed to model the MIR spectrum and the overall SED of T Cha ranging from optical to millimetre wavelengths. Specifically, we  attempted a best-fit of the JWST spectrum within $5-25$ $\mu$m, which includes the MIR thermal dust continuum emission along with the AIBs around 6.2, 8.1, and 11.3 $\mu$m, and the optical ($0.44-0.9$ $\mu$m) and millimetre ($0.85-3.3$ mm) band fluxes. Noting the `see-saw' variability of the disk, we excluded all the photometric data points within $1-500$ $\mu$m from our modelling procedure.
        
        It is trivially assumed that the IR excess in $\sim5-9$ $\mu$m continuum mostly comes from the inner disk region, while the outer disk region mainly contributes to the continuum $>14$ $\mu$m. 
        It is important to note that PAHs are assumed to be dust grains and their emission is approximated as Drude profiles, which have extended wings (Sect. \ref{sec:pahprofiles}). Hence, they contribute to the thermal continuum in the $5-14$ $\mu$m region. To find the best model-fit to the observed data, we varied all the relevant parameters including dust abundances, as well as the physical disk structure that governs the thermal emission and, thus, shapes the disk's SED. To reduce the degeneracies involved with the model, we utilise all the relevant information from the previous studies, mainly from O13, H15, P17, and H18.
        
        \subsection{Input stellar spectrum} \label{sec:starparameters}
        
        It is important to well-quantify the radiation source for the radiative transfer model. To obtain a suitable stellar spectrum, we refer to the database named The Castelli AND Kurucz 2004 Stellar Atmosphere Models\footnote{\url{https://archive.stsci.edu/hlsps/reference-atlases/cdbs/grid/ck04models/}} \citep{CastelliKurucz2003}. Given that the central star of T Cha is of G8 spectral type, we assumed a 5500 K stellar temperature and we chose the corresponding atmospheric model, with the stellar gravity given as log$(g\:\mathrm{cm}\:\mathrm{s^{-2}})=3.5$. Then we calculated the stellar spectrum with absolute fluxes considering a stellar radius of $1.3$ $R_\sun$ and used that spectrum as the input in our entire modelling process. 
        
        Since the disk of T Cha has a high inclination close to $\sim$70$^{\circ}$, as reported in literature (see Sect. \ref{sec:inclination}), some of the stellar radiation might become extinct, while transmitting through the inner disk and the outer disk surface layers towards the observer. \cite{Xie+2025} reported that the JWST observations were taken during the brightest phase of the variable photosphere of the central star. Therefore, we have aimed to match the highest photometric fluxes in the variability curve of the stellar photosphere. For this purpose, we took care to ensure that the modelled disk should not block more than $\sim$1\% of the stellar irradiation, as calculated by noting the flux around 0.5 $\mu$m in the input stellar spectrum.
        
    \subsection{Density structure} \label{sec:densityprofile}
        
        The adopted density structure is that of a disk in a vertical hydrostatic equilibrium with an axisymmetric density profile,
        \begin{equation}
                \rho(r, z) = \frac{\Sigma(r)}{\sqrt{2\pi}H(r)}\mathrm{exp} \left( -\frac{1}{2} \left[ \frac{z}{H(r)} \right]^2 \right);\\  
                \Sigma(r) = \Sigma_0 \left( \frac{r}{R_0} \right)^{\gamma}.
                \label{Eqn:5}
        \end{equation}
        $\Sigma(r)$ denotes the surface density as a function of the radius having an exponent $\gamma$, with a reference value given by $\Sigma_0$ at the reference radius, $R_0$. $\rho(z)$ has a Gaussian form with the maximum at the midplane and distribution along the vertical axis ($z$). The $1\sigma$ deviation of the midplane density along $z$ is referred to as the scale height, with its radial form expressed as         
        \begin{equation}
                H(r) = H_0 \left( \frac{r}{R_0} \right)^\beta,
                \label{eqn:6}
        \end{equation}
        where the radial exponent, $\beta$, is called the flaring index and $H_0$ is the reference scale height at $R_0$.      
        
        To calculate radiative transfer in {\tt MCFOST}, the density structure is defined over a mesh-grid that can be radially divided in multiple components. For each component, we specify inner and outer radii, total dust mass ($M_\mathrm{d}$) and scale height, in terms of $H_0$, $R_0$, and $\beta$, as input parameters. Thus, we parametrised the radial dust surface density ($\Sigma_\mathrm{d}(r)$) and scale height ($H(r)$) in the model, assuming the density form from Equation \ref{Eqn:5}.
        
        Since it is well-accepted that T Cha has a gap in the dust disk, we primarily considered two components within the density grid: an inner and an outer disk, radially separated by a gap with $\Sigma_\mathrm{d}=0$.  
        However, the ALMA CO line observations did not infer a gap in the gas disk (H15; \citealt{Wolfer+2023}). Also, spatially resolved ALMA continuum observations have so far only confirmed a gap in the larger millimetre-sized grains H18. Hence, 
        we also explored the possibility of smaller dust grains in the so-called dust gap, where we added another component within the grid representing the `gap' region with non-zero dust mass.  
        
        The inner edge of the inner disk is directly illuminated by the central star and achieves a very high temperature. This results in a puffing-up of the inner disk scale height within a thin rim at its inner edge \citep{Dullemond+2001}. Hence, we assumed the inner disk geometry in our model would have such a `puffed-up' inner rim component, with a higher scale height than the rest of the inner disk.
        
        \subsection{Disk orientation} \label{sec:inclination}
        
        The knowledge of disk orientation in the sky is important for the PPDs, since the observables may depend sensitively on the inclination ($i$) and position angle ($PA$) of the disk. Hence, it is crucial to properly set the disk orientation in the radiative transfer models to compare with the observations. In our modelling, we set $PA=90^\circ$ arbitrarily since we only fit the total spectrum from the disk and do not compare any disk images. The inclination of the disk of T Cha has been previously estimated in the range of $\sim58^\circ-75^\circ$ (\citealt{Brown+2007}{; O13; H15; P17; H18}). The inclination of 73$^\circ$ obtained by H18 from the ALMA 3.3 mm continuum image corresponds mostly to the settled dust. H15 obtained $i=67^\circ$ by fitting the ALMA CO line emission, while the modelling of the VLT-SPHERE 1.6 $\mu$m scattered light observations by P17 resulted in $i=69^\circ$ and, hence, a lower estimation of inclination (average $i=68^\circ$) is obtained from the observations tracing the disk surface.  
        Since the main scope of this work is to model the PAH emission that mainly traces the disk surfaces, we set $i=68^\circ$ for our T Cha disk model. Presuming the inclination to be well-constrained by these resolved observations of the outer disk, we chose not to vary the inclination to reduce our model degeneracies. 
        
        \subsection{Dust properties} \label{sec:dustpropeties}
        
        We considered both carbonaceous and silicate dust for the modelling. The carbonaceous dust include PAHs and amorphous carbon. Stochastic heating of PAHs are considered in the radiative transfer calculations. We adopted the opacities for amorphous carbon calculated from the optical properties given by \citet{Zubko+1996} for BE{\footnote{BE sample refers to the carbon grains produced by burning benzene in air under normal conditions \citep{Zubko+1996}}} amorphous carbon materials}. The amorphous silicate opacities considered are derived from that of the astronomical silicates described in \citet{Draine2003a, Draine2003b}. We discuss the PAH opacities used in this work in Sect. \ref{sec:pahopacities}. We considered all the dust species to follow a power-law size distribution. 
        
        While dust settling naturally occurs in disks, the amount of settling is difficult to adopt in the models due to the absence of relevant constraints (e.g. turbulence). However, completely excluding such effect in the model calculations might deviate the model parameters compared to the realistic solutions. Dust settling can considerably affect the emission from small dust, especially PAHs, even with high turbulence levels ($\alpha=0.01$) as shown by \cite{Dullemond+2007}.   
        Hence, in our model calculations, we allowed for a standard settling scheme for the dust grains according to their sizes and radial position in the disk as prescribed in \cite{Dubrulle+1995}.    
        
        \subsection{PAH opacities} \label{sec:pahopacities}
        
        Similarly to the approach taken in astrophysical dust models, we interpreted PAHs as dust grains instead of molecules, while their opacities in our radiative transfer model were incorporated in terms of dust opacities. We utilised the PAH opacities corresponding to the astro-PAH model, originally described in \cite{LiDraine2001} (also refer to \citealt{DraineLi2007} and \citealt{Draine+2021}). In particular, these are empirical opacities derived to replicate the AIBs observed mainly in the spectra of galaxies, nebulae, and ISM, in general; they were also guided by the laboratory spectra of PAHs. The opacities were further modified based on the Spitzer spectra of a large number of galaxies \citep{Smith+2007, DraineLi2007}. As a result, astro-PAH opacities account for a generalised PAH spectral template, where the 7.7 and 8.6 $\mu$m bands are well separated, mostly corresponding to class $\mathcal{A}$ and class $\mathcal{B}$ spectral types (see previous discussion in Sect. \ref{sec:pahprofiles}).
        
        Therefore, it is not possible to generate more specific class $\mathcal{C}$ signatures of a PAH profile directly using the current astro-PAH opacities. 
        A close replication of the observed positions and shapes of the AIB profiles in the model spectrum allows for a more accurate estimation of the modelled PAH characteristics.
        Hence, we calculated the PAH opacities, based on the original formalism described in \citet{LiDraine2001}, to replicate the class $\mathcal{C}$ AIB features in the JWST spectrum of T Cha. For this purpose, we used our AIB profile analyses described in Sect. \ref{sec:pahprofiles} to modify the relevant parameters that modulates the position and shape of the PAH profiles in the astro-PAH model. Details of this modification procedure are described in Appendix \ref{app:pahmodified}. We used the modified PAH `opacity files' as the input for the code to run the dust radiative transfer calculations. 
        
        We considered both neutral and ionised PAHs with opacities derived for a size range of $\sim2.88-50$ \AA. The size range correspond to PAHs with $\sim10-50000$ C atoms PAHs, referring to the expression in \cite{Draine+2021} given as        
        \begin{equation}
                N_C = 418\left(\frac{a}{10 \AA}\right)^3,
                \label{eqn:7}
        \end{equation}
        where $a$ is the radius of the assumed PAH grain and $N_\mathrm{C}$ denotes the number of C atoms in a PAH molecule. This considers a carbon mass density of $2.0$ g cm$^{-3}$ for the grain material. We note that smaller PAHs of $<20$ C atoms were generally excluded in the astro-PAH model, as they are photolytically unstable under interstellar radiation conditions \citep{LiDraine2001, Draine+2021}. However, since the environment of a T Tauri star as in the disk of T Cha is expected to have a milder UV radiation, we allowed the PAHs to be smaller (also see \citealt{Allamandola+1989} for a similar argument for the Red Rectangle). We assumed an MRN \citep{Mathis+1977} type grain size distribution (power-law index $p=-3.5$) for the PAHs. 
        
        In our model, we included PAHs only in the outer disk (see discussion in Sect. \ref{sec:morphologydust}). The quantities: PAH size range, the ionized-to-neutral PAH ratio (ionization fraction), and the ratio of the total PAH mass to the total dust mass (PAH-to-dust mass ratio) in the model are varied and constrained by fitting the AIB fluxes and the their flux ratios.  
        
        \section{Model optimization: Results and discussion} \label{sec:resultsdiscussion}
        
        We present a model of the PPD around T Cha that satisfactorily fits the observed JWST spectrum, including the dust continuum and the AIBs and also the optical and millimetre band fluxes, as shown in Fig. \ref{fig:model}.
        The main parameters defining the model are summarised in Table \ref{tab:disk}. 
        The model involves a large number of parameters, which often influence the SED interdependently. Thus, we ran several sets of model-grids and applied both $\chi^2$ minimization and visual interpretation at different modelling stage. Therefore, we attempted to narrow down the tentative parameter ranges and improve the model iteratively. This section discuss the best fiducial model obtained from this work, along with assumptions and optimization process related to specific model parameters and possible model degeneracies.    
        
        \subsection{Disk morphology and dust population} \label{sec:morphologydust}
        
        We reinstated the basic morphology of the T Cha disk as found by the earlier studies, which is of a typical transition disk with an inner and an outer disk separated by a dust gap. However, we also propose the possible presence of smaller dust grains within the so-called dust gap. Our model predicts a total dust mass of {$\sim7.6\times10^{-5}$} $M_{\sun}$ in the disk of T Cha. The distribution of different dust species among the modelled disk components are summarized in Table \ref{tab:dust} in terms of their mass and sizes. Figure \ref{fig:densityprofile} shows the radial profiles of the disk scale height and the dust surface density corresponding to our best fiducial model.   
        
        \subsubsection{Inner disk dust mass} \label{sec:innerdisk}
        
        The only model parameter  we varied for the inner disk is the dust mass, while the rest of the inner disk parameters were kept fixed to the values guided by the literature, {mostly from O13,} which considerably reduce the degeneracies in our model. We took this approach since the main scope of this work was to address the PAH emission from T Cha, which presumably arise from outer disk region, as discussed later. 
        
        In order to fit their VLTI interferometric data, O13 found a very thin inner disk located at $0.07$ au from the central star and extended only up to $0.11$ au. However, they further proposed a `hidded-mass', in the shadow of the optically thick inner disk, might be present and may extend up to 5 au. From subsequent ALMA observations, H18 reported an unresolved inner disk with an extent $<1$ au in thermal emission from mm size dust. Based on these results, we parametrized the inner disk in terms of two regions: a thin `puffed-up rim', located at the inner disk edge, having an extent of $0.07-0.11$ au with a scale height of $0.2$; then there is the rest of the inner disk with a lower scale height of 0.05 au at 1 au, extending from 0.11 to 1 au, which we refer to as the `shadowed region'. The flaring index of the inner disk ($0.07-1$ au) was set to 1.0, also following O13.   
        
        We considered a mix of amorphous carbon and silicate grains in a mass ratio of $1:3$ for the inner disk. The absence of the 10 $\mu$m silicate emission band in the Spitzer spectra of T Cha was previously interpreted as the absence of the small silicate grains in the inner disk. O13 found a lower limit of 5 $\mu$m for the size of the silicate grains in their modelling. We directly adopted the size range  of $5-1000$ $\mu$m for the silicate grains and as $0.01-1000$ $\mu$m  for that of amorphous carbon grains  in our model from O13. Both the grain population follow an MRN type power-law size distribution with a slope of $-3.5$.  We did not include PAHs in the inner disk since O13 found no PAH emission from the inner disk in their MIR interferometric data. Also, from the analyses of the JWST MIRI-IFU slices by \cite{Bajaj+2024}, the PAH emission is more extended than the continuum and seems to trace the outer regions of the disk. 
        
        We reduced the dust mass in the inner disk compared to that in O13 in order to match the $5-9$ $\mu$m continuum in the JWST spectrum, which is a factor of $\sim$3 lower than that in the Spitzer spectrum. The reduction of the dust mass in the inner disk results in an effective lowering of the scale height, due to a decrease in the opacity and, thus, it increased the outer disk irradiation, which was expected to help interpret the observed `see-saw' variability. We note that the puffed-up rim may no longer be optically thick, due to its reduced dust mass, and, hence, the shadowed region might contribute considerably to the visible thermal emission in $5-9$ $\mu$m. Hence, we varied the dust mass in both the puffed-up rim and the shadowed region. We find that dust masses of $1.2\times10^{-12}$ $M_\sun$ within the puffed-up rim and of $1.9\times10^{-11}$ $M_\sun$ in the shadowed region of the inner disk accounts well for the 5-9 $\mu$m JWST continuum. 
        
        \subsubsection{Outer disk geometry and dust distribution} \label{sec:outerdisk}
        
        We presume that the `see-saw' variability would also affect the far-IR (FIR) Herschel photometric fluxes in the range of $70-500$ $\mu$m. However, we may also assume that the variability would decrease with increasing wavelength since the larger dust must be more settled and less affected by the increased radiation on the outer disk. Hence, we used the MIR continuum in the $14-25$ $\mu$m range and the millimetre fluxes at ALMA $0.85$ and $3$ mm, SEST $1.3$ mm, and ATCA $3.3$ mm bands
to constrain the outer disk characteristics. 
        
        Our approach to fit the $14$ $\mu$m to 3.3 mm SED is to match the slope of the observed SED in $14-25$ $\mu$m and $0.85-3.3$ mm regions, while simultaneously reproducing the observed peak strength of the continuum around $\sim$25 $\mu$m. We included amorphous carbon and silicate dust with a fixed standard ratio of $1:3$ in the outer disk. Variations among the amorphous carbon-to-silicate mass ratio does not have any major physical consequences to the scope of this work. PAHs are included in the outer disk and will be discussed in detail in Sect. \ref{sec:pahcharacteristics}. We also note that the inclusion of PAHs does not noticeably vary the modelling of the SED range focussed in this section. 
        
        We note that larger grains close to the orders of centimetres and a shallower size distribution than the MRN law is needed to better fit the slope of the millimetre fluxes $>1$ mm. We set a minimum grain size of 0.01 $\mu$m, while varying the maximum sizes range from $\sim1-10$ mm, and the slope ($p$) of the power-law grain size distribution from $-2.5$ to $-3.9$. The rest of the outer disk parameters set for running these models were chosen from a standard range discussed below. We obtained a good match to the slope of the $0.85-3.3$ mm SED, with dust grains of sizes $0.01-5000$ $\mu$m with $p=-3.3$.
        
        The outer disk geometry, including its radial extent and the scale height, along with the total dust mass is constrained mainly by matching the strength and slope of the $14-25$ $\mu$m continuum. The outer disk of T Cha was resolved in the 3.3 mm ALMA image by H18. We truncated the outer disk at 50 au, which matches the radius where the millimetre dust emission from the outer disk falls to a value of $\sim$10\% of the peak intensity, based on the ALMA image. The outer edge of the proposed dust gap (i.e. the inner radius of the outer disk) has been broadly varied in earlier works, across a range of $\sim12-30$ au. Hence, we varied the inner edge of the outer disk in a wide range up to 32 au. We varied the scale height ($H_0/R_0$) of the outer disk within a range of $1.5$ to $3$ au calculated at $R_0=25$ au, while the flaring index $\beta$ is varied from 1.1 to 1.5. 
        
        The outer disk dust mass also needed to be varied alongside the geometrical parameters to match the $14-25$ $\mu$m continuum strength. However, the peak flux around $\sim$25 $\mu$m ends up overestimated if we try to generate the strength of the $14-25$ $\mu$m continuum by increasing the dust mass. We also note that including a single dust population of $0.01-5000$ $\mu$m in the entire outer disk generates a flatter slope in the $14-25$ $\mu$m model SED than in the observed. Therefore, we propose the outer disk to consist of two regions. First, we have a region with smaller dust population with only up to micron-sized grains, located at the inner part of the outer disk, accounting well for the steeper rise of $14-25$ $\mu$m continuum. The rest of the outer disk region accommodates the wider grain population up to the millimetre-sized dust. This results in a much better simultaneous fit in qualitative terms to the observed $14-25$ $\mu$m continuum slope and the SED peak around $\sim25$ $\mu$m. We refer to the smaller dust region as outer disk `wall', where the inner edge of the wall is understood as the inner edge of the outer disk in our discussion. The rest of the outer disk is referred as the `outer region'. The wall and the outer region is assumed to have the same scale height. 
        
        From modelling of the 1.6 $\mu$m scattered light observations, P17 estimated the outer disk at $\sim$30 au. While the modelling of 3 mm ALMA observations by P17 revealed a peak radial intensity of the outer disk at $\sim$35 au. Therefore, the two zones in the our model with two dust populations are also justified from separation of $\sim$5 au between the larger and the smaller dust distribution reported from those observations. 
         
        Hence, we assumed a wall width of 5 au for simplicity. Furthermore, changing of the wall width would become degenerate when varying the dust mass within the wall and considering the opacity of the medium. We kept the minimum grain size in the wall at 0.01 $\mu$m and varied the maximum grain size from $0.1-10$ $\mu$m. The grain size distribution has a power-law index $p=-3.5$. We explored a wide range of dust mass for the wall in a qualitative fitting approach. For the best model calculations, we vary the dust mass in the wall in a range of $1\times10^{-7}-8\times10^{-7}$ $M_\sun$. Similarly, governed by a visual fit to the millimetre fluxes and literature ranges (e.g. O13, P17), the dust mass in the outer region is tuned between $1\times10^{-5}$ to $5\times10^{-4}$ $M_\sun$ for the best model calculations.   
        
        Finally, from a $\chi^2$ minimization of the fit to the $14-25$ $\mu$m continuum range, we estimated the outer disk of T Cha to have a radial extent of $31-50$ au, with a scale height measuring $1.9$ au at 25 au. The location of the wall was at $31-36$ au. We found a maximum grain size of 4 $\mu$m in the wall and estimate dust masses of $7\times10^{-7}$ and $7.5\times10^{-4}$ $M_{\sun}$ in the wall and outer region of the outer disk, respectively. 
        
        We found the inner radius of the outer disk from our independent modelling is quite similar to that predicted from resolved outer disk observations by P17 and H18.
        Therefore, the gap width of $\sim$30 au in our model also further supports the possible presence of a planet of $\sim1.2$ $M_\mathrm{Jup}$ or the possibility of multiple planets within the gap, as proposed earlier (\citealt{Huelamo+2011}; P17, H18).       
        \begin{figure}
                \centering
                \includegraphics[width=\columnwidth]{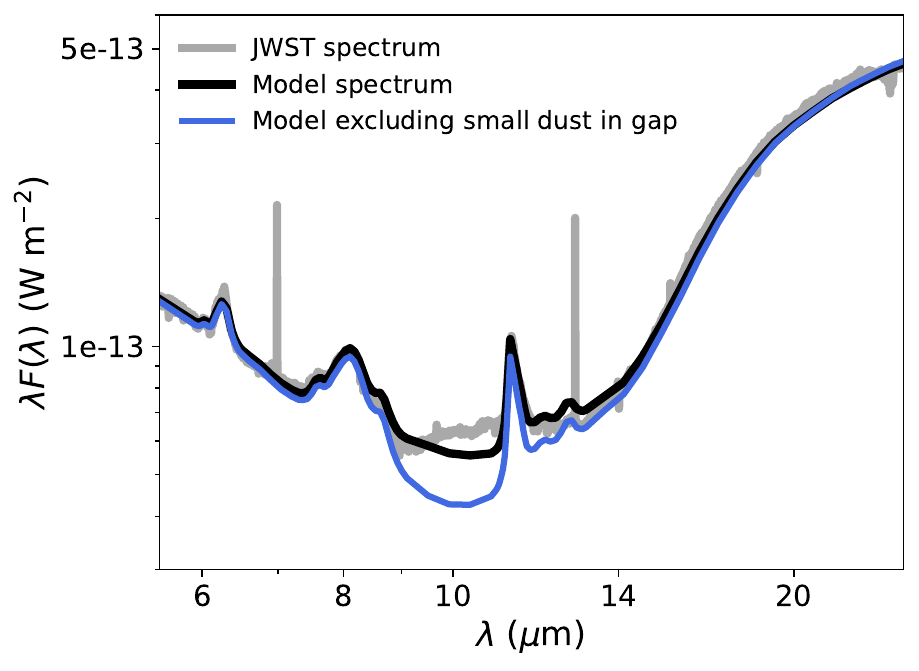}
                \caption{Comparison of the T Cha disk model computed in this paper, which includes small dust of $0.01-1$ $\mu$m within the gap region and the same model without any dust included in the gap.}
                \label{fig:gap}
        \end{figure}
        
        \subsubsection{Possibility of small dust in the gap} \label{sec:10mplateau}
        
        The JWST spectrum of T Cha clearly depicts a plateau-like shape of the continuum around $\sim$10 $\mu$m, as compared to a sharper dip in case of the Spitzer spectra. The presence of smaller dust grains within the gap of T Cha was not interpreted during the modelling of the Spitzer observations (e.g. \citealt{Brown+2007, Cieza+2011}; O13). The regular PAH spectral templates depict a plateau between the 8.6 and 11.3 $\mu$m bands. However, the observed JWST continuum around $\sim$10 $\mu$m in is much shallower compared to that generally found between those bands. Furthermore, the ALMA CO line observations have not so far reported any gap in the gas disk (H15; \citealt{Wolfer+2023}). Some proof of the presence of dust in the gap could support the possibility of replenishment of the inner disk after its destruction as suggested by \cite{Xie+2025}. 
        
        Therefore, we propose a faint 10 $\mu$m silicate emission arising from the gap that compensates the otherwise deeper plateau within the 8.6 and 11.3 $\mu$m bands, which could explain the JWST continuum plateau around $\sim$10 $\mu$m. We assumed smaller dust grains that might be coupled with the accreting gas and, hence, we filled the dust gap formed in terms of the larger millimetre-sized grains. Thus, in our models, we simply considered sub-micron amorphous carbon and silicate dust grains in a $1:3$ ratio, having a size range of $0.01-1$ $\mu$m with an MRN type size-distribution, within the gap spanning $1-31$ au. We defined the density distribution of matter within the gap roughly following \cite{Maaskant+2014}. The scale height was kept at $H_0/R_0=0.05$ defined at 1 au, the same as that for the shadowed region of the inner disk. We assumed a flaring of the less dense matter within the gap with $\beta=1.1$. The surface density profile is considered to be flat with $\gamma=0$. By exploring a range of dust mass, we found that a dust mass of $6\times10^{-10}$ $M_{\sun}$ considered within the gap can closely match the strength of the continuum around $\sim$10 $\mu$m. Figure \ref{fig:gap} clearly shows an improvement of the overall model-fit to the JWST spectrum considering the abundance of small dust within the gap.       
        
        We did not include PAHs in the gap as well, since we note that the dust mass about five orders lower in the gap, adding PAHs with same PAH-to-dust ratio as found in our model (see Sect. \ref{sec:pahcharacteristics}) does not make any noticeable difference in the modelled PAH fluxes. A prevalence of ionised PAHs in the gaps of PPDs were earlier predicted by \cite{Maaskant+2014} considering a higher PAH-to-dust mass ratio equalling that of the ISM. However, that is still about four orders lower than that of the outer disk and it still does not affect the main conclusions of this paper. A higher gas-to-dust ratio in the gap could have enhanced the PAH emission, although such observations have not been  available thus far for T Cha and could be investigated in the future.
        
        \begin{figure}
                \centering
                \includegraphics[width=\columnwidth]{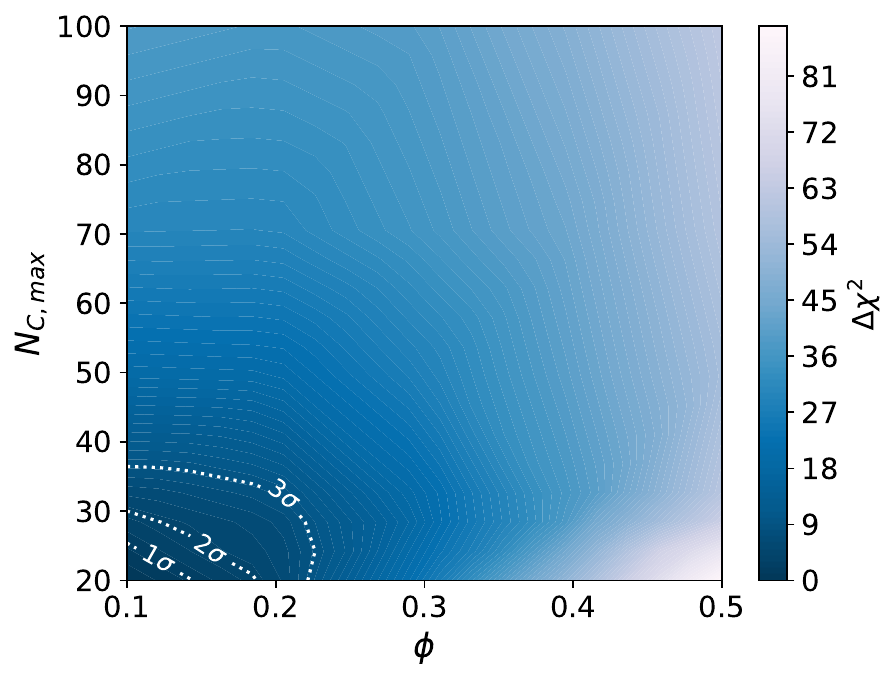}
                \caption{$\Delta\chi^2$ contours of the fits to the observed 6.2/8.1, 6.2/11.3, and 8.1/11.3 $\mu$m AIB ratios in $N_\mathrm{C, max}-\phi$ parameter space (for $N_\mathrm{C, min}=10$ and $f_\mathrm{PAH}=20\%$ corresponding to the minimum $\chi^2$ model, see text in Sect. \ref{sec:pahsizechargemass}). The $\sigma$ levels are indicated with dotted contour lines. The $1\sigma$ contour helps in predicting the fiducial model values: $N_\mathrm{C}=10-26$ and $\phi=0.15$, thereby estimating the final PAH-to-dust mass ratio as $f_\mathrm{PAH}=17\%$ for the best fiducial model (Fig. \ref{fig:fpaherror} and Sect. \ref{sec:pahsizechargemass}).}
                \label{fig:pahsizecharge}
        \end{figure}
        
        \begin{figure}
                \centering
                \includegraphics[width=\columnwidth]{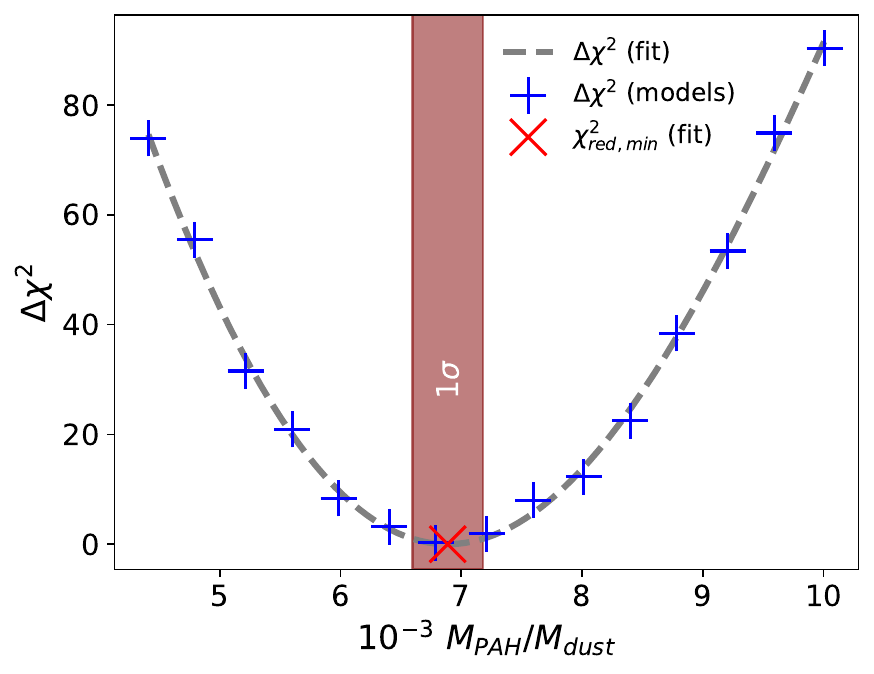}
                \caption{Plot of the $\Delta\chi^2$ values of the model fits with different PAH-to-dust mass ratios ($M_\mathrm{PAH}/M_\mathrm{dust}$), while the rest of the model parameters are fixed to the best fiducial model, with optimised PAH size range ($N_\mathrm{C}$=10-26) and ionization fraction ($\phi$=0.15), as discussed in Sect. \ref{sec:pahsizechargemass}, and that best fits the continuum (Sect. \ref{sec:morphologydust}). The spectral regions covering the entire AIB profiles are used for the $\chi^2$ calculations. The minimum of the plot ($\Delta\chi^2=0$)
                corresponds to the minima of the fit through the $\chi^2$ values (also see text in Sect. \ref{sec:pahsizechargemass}). Hence, the light red shaded region with $0\leq\Delta\chi^2\leq1$, shows the $1\sigma$ confidence limit in the estimation of $M_\mathrm{PAH}/M_\mathrm{dust}$, given the fiducial $N_\mathrm{C}$ and $\phi$ values.}
                \label{fig:fpaherror}
        \end{figure}
        
        \subsection{PAH abundances} \label{sec:pahcharacteristics}
        
        From laboratory and observational studies, it has been found that the PAH emission bands are sensitive to PAH sizes (expressed here in terms of the number of C atoms, $N_\mathrm{C}$) and ionization fraction ($\phi$) of the PAH population (e.g. \citealt{Tielens2008}). Emission from the smaller PAHs is stronger in shorter wavelength bands. Hence, the 6.2/7.7, 6.2/11.3, and 7.7/11.3 $\mu$m band ratios increase for a lower $N_\mathrm{C}$. The 6.2 and 7.7 $\mu$m band emissions increase with a higher fraction of ionised PAHs (e.g. \citealt{Allamandola+1999}). Hence, the 6.2/11.3 and 7.7/11.3 $\mu$m band ratios are proportional to $\phi$. Further, the 6.2/7.7 $\mu$m band ratio from a purely ionised to a purely neutral population of PAHs increases by $\sim$10\%{\footnote{The reader may refer to Fig. 21 in \cite{Draine+2021} for a visual summary of the behaviour of the band ratios with varying PAH sizes and ionisation fraction.}} (e.g. \citealt{Draine+2021}). 
                
        Therefore, the observation of multiple prominent AIBs in the JWST spectrum of T Cha is helpful in predicting $N_\mathrm{C}$ and $\phi$ in the disk through our modelling. We also estimated the PAH mass, expressed in terms of the PAH-to-dust mass ratio ($M_\mathrm{PAH}/M_\mathrm{dust}$) in the disk of T Cha by tuning the PAH mass in our model. We note that the estimation of the total PAH mass by fitting the absolute AIB fluxes is more accurate when the band ratios are closely reproduced in the modelled spectrum.    
        
        \subsubsection{Exploration of parameters} \label{sec:pahabundancegrid}
        
        We primarily aim to explore different possible $N_\mathrm{C}$ and $\phi$ values in order to obtain the best fit to the 6.2/8.1, 6.2/11.3, and 8.1/11.3 $\mu$m AIB ratios. Since our modified PAH opacities are based on the astro-PAH model, we assumed the 8.1 $\mu$m feature depends on $N_\mathrm{C}$ and $\phi$ in a similar way as in case of the 7.7 $\mu$m band. The band-to-continuum ratio is dependent on the $M_\mathrm{PAH}/M_\mathrm{dust}$ in the medium. Since the PAH emission is generated from the disk, which is not fully optically thin in this case, the total modelled PAH flux is not always linearly proportional to the input PAH mass. In other words, the best fitting $N_\mathrm{C}$ and $\phi$ is also a function of the chosen $M_\mathrm{PAH}/M_\mathrm{dust}$. Hence, we also varied the $M_\mathrm{PAH}/M_\mathrm{dust}$ alongside $N_\mathrm{C}$ and $\phi$, and simultaneously fit the AIB fluxes and the band ratios for a proper estimation of the PAH abundances.  
         
        From our initial effort to fit the AIBs with the modelled PAH bands, we predicted $M_\mathrm{PAH}/M_\mathrm{dust}>4\times10^{-3}$ $M_\sun$ by matching the flux of the 8.1 $\mu$m AIB. This amounts to $f_\mathrm{PAH}>10\%$, where $f_\mathrm{PAH}$ stands for the fraction of $M_\mathrm{PAH}/M_\mathrm{dust}$ in the disk compared to the value of $M_\mathrm{PAH}/M_\mathrm{dust}\simeq4\times10^{-2}$ corresponding to the ISM according to \cite{DraineLi2007}. We then explore wide parameter spaces of the minimum ($N_\mathrm{C, min}$) and maximum ($N_\mathrm{C, max}$) probable PAH size ranges (Sect. \ref{sec:pahopacities}, $p=-3.5$) and $\phi$, while setting $f_\mathrm{PAH}$ at 10\%. We find $N_\mathrm{C, min}$ converging to the lowest available size of 10 C atoms, while $N_\mathrm{C, max}<100$ and $\phi<0.5$.
 
        To obtain a more robust prediction of the parameters, we then ran a grid of models using $N_\mathrm{C, min}=10$ and $N_\mathrm{C, max}=20$, 30, 50, 70, and 100; $\phi=0.1-0.5$, in an interval of 0.1; and $f_\mathrm{PAH}=10-60\%$, with an interval of 10\%.       
                        
    \subsubsection{Predicting PAH sizes, ionisation fraction, and mass} \label{sec:pahsizechargemass}
        
        For individual models from the grid defined above, we calculated the $\chi^2$ values of the fits as 
    \begin{equation}
        \chi^2 = \sum_{n=1}^{N}\frac{\left( \mathcal{O}_n-\mathcal{M}_n\right)^2}{\sigma_n^2},
        \label{eqn:8} 
    \end{equation}
    where $\mathcal{O}_\mathrm{n}$ and $\mathcal{M}_\mathrm{n}$ stand for the observed and modelled values of the $n$th observable, respectively. The error in the $n$th observable is given by $\sigma_\mathrm{n}$. $N$ denotes the number of observables; in this case, there are six, including peak-to-peak flux values (see Appendix \ref{app:peaktopeakflux}) of the three AIBs: 6.2, 8.1, and 11.3 $\mu$m, and the three band ratios: 6.2/8.1, 6.2/11.3, and 8.1/11.3 $\mu$m. We find that the minimum $\chi^2$ value corresponds to the model ($\chi^2_\mathrm{min}$-model) with $N_\mathrm{C, min}=10$, $N_\mathrm{C, max}=20$, $\phi=0.1$, and $M_\mathrm{PAH}/M_\mathrm{dust}=8\times10^{-3}$ ($f_\mathrm{PAH}=20\%$).  

    Hence, the values of $N_\mathrm{C, max}$ and $\phi$ from the $\chi^2_\mathrm{min}$-model points to the lowest values of these parameters chosen for the grid. As we continue to explore lower ranges of $N_\mathrm{C, max}$ and $\phi$, we note that $N_\mathrm{C, max}$ is indeed unbound in terms of the lower limit, and also that the model then requires an unusually high $M_\mathrm{PAH}/M_\mathrm{dust}$ ($f_\mathrm{PAH}\sim50\%$), about twice of the maximum $f_\mathrm{PAH}$ found among the disks by \cite{Woitke+2019}. This behaviour is mainly due to a strong 6.2 $\mu$m band in T Cha, which theoretically requires the smallest available PAH population for a best model fit. However, such results may not be physically correct owing to a few reasons described in the following.
   
    We point out that the band ratios may also depend on some other factors. For example, dehydrogenation of PAH molecules increases the 6.2 and 7.7 $\mu$m band strengths since they are associated with the C-C vibrational modes of the PAHs \citep{Schutte+1993, Andrews+2016PAH}. The anharmonicity present in the PAH molecular vibration modes affects the band profile shapes \citep{Mackie+2016}. 
    Furthermore, the aliphatic-to-aromatic ratio (discussed in Sect. \ref{sec:pahprofiles}) could modulate the exact dependencies of the band ratios on $N_\mathrm{C}$ and $\phi$. The empirically derived astro-PAH opacities assumed for this work do not fully consider these effects. Finally, a precise estimation of $N_\mathrm{C}$ and $\phi$ also requires an accurate fit to the dust continuum, which we could not achieve in the $9-14$ $\mu$m continuum range in this work. 
      
    To obtain the final values of $N_\mathrm{C, max}$ and $\phi$ for our best fiducial model, we further calculated $\Delta\chi^2 = \chi^2 - \chi^2_\mathrm{min}$, which gives the uncertainties around the $\chi^2_\mathrm{min}$-model parameters defining the PAH abundances. For this, we needed to rescale the observational errors ($\sigma$) such that the reduced $\chi^2$ ($\chi^2_\mathrm{red}$){\footnote{{$\chi^2_\mathrm{red}=\chi^2/k$, where, $k$ is the degrees of freedom of the fit, given by $k=N-p$, while $p$ is the number of varied parameters.}}} value of $\chi^2_\mathrm{min}$-model becomes unity. The $\Delta\chi^2$ contours for the model fits with different $N_\mathrm{C, max}$ and $\phi$, with $f_\mathrm{PAH}=20\%$ (corresponding to the $\chi^2_\mathrm{min}$-model) are shown in Fig \ref{fig:pahsizecharge}. Assuming two free parameters, $N_\mathrm{C, max}$ and $\phi$, we also overplot the $\sigma$ levels, which gives approximate uncertainty ranges in the modelled $N_\mathrm{C, max}$ and $\phi$. Similar plots showing $\Delta\chi^2$ contours for the $N_\mathrm{C, max}-f_\mathrm{PAH}$ and the $\phi-f_\mathrm{PAH}$ parameter spaces are shown in Fig. \ref{fig:pahsizemass} and Fig. \ref{fig:pahchargemass}, respectively. 
    
    Using the $1\sigma$ level of the $\Delta\chi^2$ contours for the $N_\mathrm{C, max}-\phi$ space and considering that other factors might influence the band ratios as discussed above, we predicted $N_\mathrm{C, max}\simeq26$ and $\phi\simeq0.15$. Hence, we estimated a PAH size distribution of $N_\mathrm{C}=10-26$ and $\phi=0.15$ for the best fiducial model of T Cha. Given these values, we attempted to infer on the estimate of $M_\mathrm{PAH}/M_\mathrm{dust}$. For this, we ran a further set of models by varying $M_\mathrm{PAH}/M_\mathrm{dust}$ around the $\chi^2_\mathrm{min}$ model value, in an interval of $f_\mathrm{PAH}=1\%$. For these models, we calculated $\chi^2$ values of the fits to the entire AIB profiles. We rescaled  $\chi^2$ so that the reduced $\chi^2$ would become unity, by tuning the observational errors in the JWST spectrum. We then fit a polynomial function through the $\chi^2$ values and  measured $\Delta\chi^2$, shown as a function of $M_\mathrm{PAH}/M_\mathrm{dust}$ in Fig. \ref{fig:fpaherror}. We can define the $1\sigma$ limit around the best-fit value given by $\Delta\chi^2=1$ and, thus, we have the error in the calculation of $M_\mathrm{PAH}/M_\mathrm{dust}$. From this exercise, we can obtain $M_\mathrm{PAH}/M_\mathrm{dust}=6.89(\pm0.29)\times10^{-3}$ that corresponds to $f_\mathrm{PAH}=17.2(\pm0.73)\%$.

    \subsubsection{Implications of the estimated PAH abundances} \label{sec:pahimplications}    
        
        From our modelling, we finally predict that the disk of T Cha has a population of smaller PAHs of $\sim10-26$ C atoms, with $\sim$85\% neutral PAHs, and a PAH-to-dust mass ratio amounting to $\sim$17\% of the ISM ratio of PAH-to-dust mass. Hence, our estimated PAH-to-dust mass ratio is consistent with earlier predictions while points to a high value among disks. Our prediction of small PAH sizes and low ionization fraction largely follows the results from a few previous studies. The analyses of AIBs using the laboratory and theoretical PAH database suggested that the fitting of class $\mathcal{C}$ profiles might need a random distribution of smaller ($N_\mathrm{C}<30$) and larger ($N_\mathrm{C}\geq30$) PAHs, as well as a higher fraction of neutral PAHs \citep{Cami2011, Bauschlicher+2018}. \cite{Arun2025} also predicted a population of smaller PAHs with $N_\mathrm{C}<30$ in T Cha comparing to a theoretical size-charge grid for PAHs computed by \cite{Maragkoudakis+2020}. In the context of PAH abundance uncertainties due to their cluster formation and adsorption into the larger grains, \cite{Lange+2023} theoretically predicted that PAHs could be re-processed by UV irradiation as the larger grains reach the disk surface due to vertical mixing within a turbulent disk. A prevalence of smaller PAHs, with $N_C\leq54$, towards the disk surface has been predicted as they are easier to get liberated due to such re-processing \citep{Lange+2021, Lange+2023}.    
        
        Theoretical models have shown that the presence of PAHs on the disk surfaces can be very important in driving FUV phtoevaporative wind and limiting the gas lifetime in the disks \citep{Gorti+2009, GortiHollenbach2009}. \citet{Nakatani+2023} showed that if PAH mass is depleted to $\sim0.1-1\%$ of that in the ISM, the disk would evolve into a gas rich debris disk due to the lack of FUV photoevaporation. However, a higher PAH abundance larger than this critical value can drive H$_2$ wind driven by FUV photons and reach a mass loss rate up to $\sim10^{-9}-10^{-8}$ $M_{\sun}$ yr$^{-1}$, which is close to the X-ray photoevaporation rate and much higher than the EUV photoevaporation rate of $\sim10^{-10}$ $M_{\sun}$ yr$^{-1}$ (e.g. \citealt{Gorti+2009}). The higher PAH-to-dust mass ratio in T Cha might be indicative of a high efficiency of FUV photoevaporation active on the disk surface.      
        
        \subsection{Model evaluation} \label{sec:modelevaluation}
        
        Our model is not physically unique, since the explored parameter space might be limited due to generalised assumptions and the model may have certain degeneracies if explored in further detail. However, based on the literature, we have been able to reduce the degeneracies in this particular work. 
        
        The emission from the PAHs depend on the opacities of the other dust species within the environment in a dust radiative transfer calculation. Therefore, a properly reproduced thermal continuum ensures a proper estimation of the PAH abundances. From Fig \ref{fig:pah}, we can understand the contribution of the PAHs over the thermal continuum produced by the other dust species. Our model satisfactorily reproduce the JWST continuum, except the $9-14$ $\mu$m region. The PAH abundances derived in this work mainly depend on the proper thermal balance in the outer disk, which may have been reasonably achieved, noting the fit of the $14-25$ $\mu$m continuum and the millimetre band fluxes. We further note that our modelled values of the outer disk parameters are in good agreement with the previous studies, which performed robust analyses of the resolved outer disk data (e.g. H15, P17, H18).   
        
        By exploring the models with minor deviation from our best fiducial model, we find that some degeneracies may exist between, for example, the scale height and the dust mass of the wall, as well as the flaring index $\beta$. We note that the variation in the flaring index $\beta$ also simultaneously changes the effective scale height at the inner edge of the outer disk. To observe the dependency of the model on a particular outer disk parameter, we have shown the variation in the model spectrum by changing one parameter of the outer disk at a time in Appendix \ref{app:modelset}. In Fig. \ref{fig:modelsetbeta}, we explore the $\beta$ dependency of the model (also adjusting the respective scale heights, see figure caption) to find that there is no significant difference among the model fit to the optical, MIR and millimetre continua. Differences are visible in the PAH band strength and in the FIR continuum within $30-500$ $\mu$m that was mainly covered by Herschel and which is supposed to be affected by the `see-saw' variability and, hence, we could not fit it.

        Therefore, an important source of uncertainty in estimating the PAH-to-dust mass ratio might be the degeneracy affected modelled value of the flaring index, which influences the PAH emission. However, we find the $\beta$ values from the previous studies to be within $1.0-1.1$, particularly the results from the observations in gas emission and scattered light (e.g. H15, P17) that traces the disk surface well. Hence, we expect that our estimation of $\beta=1.1$ is very close to the actual value and, hence, may rely on the PAH-to-dust mass ratio estimated considering that value.

        \section{Summary and conclusions} \label{sec:summaryconclusions}
        
        We present a parametric dust radiative transfer model of the disk around T Cha that reproduces the MIR JWST spectrum of the object obtained using MIRI-MRS instrument. Our fitting of the JWST spectrum re-establishes the geometric structure of the disk: an inner and an outer disk separated by a dusk gap, as inferred from earlier Spitzer, VLT, and ALMA observations, while also proposes the possible presence of smaller dust grains within the gap region. We satisfactorily fit the observed AIBs around 6.2, 8.1, and 11.3 $\mu$m using our modelled PAH bands generated with our best fiducial model spectrum. The prominent AIBs in the JWST spectrum allows an optimised estimation of the PAH abundances in terms of their sizes, ionisation fraction, and mass from our modelling.  
        Our main conclusions are as follows:
        
        \begin{enumerate}
                \item[-] We predicted a population of small PAHs of $\lesssim26$ C atoms in the disk of T Cha. The population consists of $\sim85\%$ neutral PAHs. Since the disk PAH emission is mainly observed from the surface layers, the PAH population more likely represent that of the disk surface. This is in line with theory of reprocessing of the clustered PAHs by the FUV irradiation that liberates only the smaller PAHs when the clusters arrive at the disk surfaces due to turbulence \citep{Lange+2023}.
                \item[-] We estimated a PAH-to-dust mass ratio of $\sim7\times10^{-3}$ in the disk of T Cha, which corresponds to $\sim$17\% of the ISM value of the PAH-to-dust mass ratio. This value is towards the higher end found among disks. Such a high PAH mass, coupled with small PAH population in the disk surface may suggest that FUV photoevaporation is efficient in case of T Cha. This might drive a mass-loss close to the order of $\sim10^{-9}-10^{-8}$ $M_{\sun}$ yr$^{-1}$ as calculated theoretically (e.g. \citealt{Nakatani+2023}). 
                \item[-] We predicted the presence of an outer disk wall having smaller dust grains of $0.01-4$ $\mu$m to better reproduce the $14-25$ $\mu$m JWST spectral continuum. In our model, the outer region of the outer disk comprise grains of sizes 0.01 $\mu$m to 5 mm, with a shallower size-distribution (power-law index $-3.3$) that better fits the SED slope in the millimeter range. This could indicate an efficient grain growth in the outer region of the outer disk of T Cha.   
                \item[-] We propose the possibility of sub-micron dust grains within the disk gap to justify the JWST continuum plateau around $\sim$10 $\mu$m. The small grains may be coupled with the accreting gas within the gap. This also supports the possibility of replenishment of the inner disk after its destruction as suggested by \cite{Xie+2025}.
        \end{enumerate}
        
        Our model is not unique owing to the probable degeneracies, which might be improved by further constraining the models using future observations. Simultaneous line and dust emission models could potentially reduce the degeneracies, which calls for a more complex treatment and this subject could be addressed in future works. 
        
        \begin{acknowledgements}
                We thank the anonymous referee for insightful suggestions that helped to improve this paper. R.B. acknowledges the funding by Agencia Nacional de Investigaci\'on y Desarrollo de Chile (ANID) - Millennium Science Initiative Program - Center Code NCN2021\_080, and also by ANID - FONDECYT Postdoctoral grant 3250171. S.C. acknowledges the support from ANID - FONDECYT Regular grant 1251456 and ANID project Data Observatory Foundation DO210001. This work is based on archival data from the observations made with the NASA/ESA/CSA James Webb Space Telescope. The original data are present in the Mikulski Archive for Space Telescopes at the Space Telescope Science Institute, which is operated by the Association of Universities for Research in Astronomy, Inc., under NASA contract NAS 5-03127 for JWST. This work shows archival data obtained with the Spitzer Space Telescope, which was operated by the Jet Propulsion Laboratory, California Institute of Technology under a contract with NASA. This research has made use of the VizieR catalogue access tool, CDS, Strasbourg, France.
        \end{acknowledgements}
        
        \bibliographystyle{aa}
        \bibliography{PPDisks}
        
        \begin{appendix}
                
                \section{Additional details on the AIB profile fitting procedure in Sect. \ref{sec:pahprofiles}} \label{app:nautilus}
                
                \cite{Smith+2007} obtained individual fits to each observed galaxy spectrum using a model, assumed to be the sum of the stellar spectrum, modified blackbody SEDs representing the dust continuum, and Drude profiles. However, we note that a full spectral fit in case of the MIR spectrum of T Cha, similar to that in \cite{Smith+2007}, is relatively complicated due to the presence of the dip and the plateau around 10 $\mu$m. Since our main scope in this paper is to use the AIB profile components to derive the modified PAH opacities (see Sect. \ref{app:pahmodified}), we assume that can be achieved from individual fits to the PAH profiles at 6.2, 8.1, and 11.3 $\mu$m as well. 
                
                First, we define the stellar flux using a fixed stellar spectrum of a star of $T_\mathrm{eff}=5500$ K, with log$(g\:\mathrm{cm}\:\mathrm{s^{-2}})=3.5$ \citep{CastelliKurucz2003}. Next, we define the modified black body continuum that will represent the thermal dust emission. In general, the inner disk region consists of hot dust, with $T\sim1000$ K, accounting for the near-IR (NIR) excess, and the rest of the inner disk SED contribution is due to colder dust of $T=100-300$ K. Thus, we make a reasonable assumption that the dust continuum under both the 6.2 and 8.1 $\mu$m AIBs can be represented by the combination of modified blackbodies of $T=1000$, 300, and 100 K. For the continuum under the 11.3 $\mu$m band, we separately define a single modified blackbody of $T=235$ K, by visually inspecting from a set of $T=200-300$ K modified blackbodies. The flux contribution (strength) of each black body component is obtained during the optimization procedure discussed below. 
                
                We performed a two-step optimization process to obtain the best-fit profile components for each AIB. Here, we describe the example of 6.2 $\mu$m band for clarity. In first step, we assume a single Drude profile representing the observed 6.2 $\mu$m AIB profile. We fit the combined spectrum of the star $+$ modified black bodies (three in case of 6.2 $\mu$m band) $+$ the single Drude profile, to the observed JWST spectrum within the region covering the 6.2 $\mu$m band, in absolute flux units. We optimise the Drude profile parameters: $\lambda_\mathrm{1}$, $b_\mathrm{1}$, and $\gamma_\mathrm{1}$ of that single profile and the fluxes of each black body in this step. We note that this provides a good approximation of the relative contribution of the wings the AIB profile and the dust continuum emission. The second optimization step assumes multiple Drude profiles for the 6.2 $\mu$m band, with varying profile parameters $\lambda_\mathrm{j}$, $b_\mathrm{j}$, and $\gamma_\mathrm{j}$, with fixed stellar spectrum, as well as fixing the fluxes strengths of the combined modified black bodies obtained in the first step. Similar approaches are taken to obtain the profile components of the 8.1 and 11.3 $\mu$m bands.
                
            \FloatBarrier
                
                \section{Modifications of the PAH opacities to replicate the AIB profiles in the JWST spectrum of T Cha} \label{app:pahmodified}
                
                \begin{figure}
                        \centering
                        \includegraphics[width=\columnwidth]{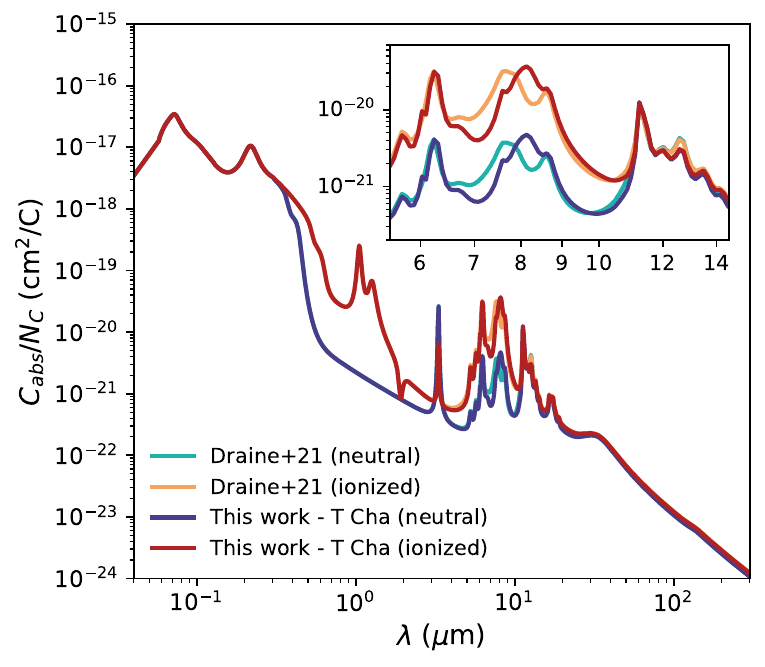}
                        \caption{Neutral and ionised PAH absorption cross section per C atom derived from  \cite{Draine+2021} and that calculated in this work for a PAH grain corresponding to {$N_\mathrm{C}=24$}.}
                        \label{fig:cabs}
                \end{figure}
                
                \begin{table}
                        \centering
                        \caption{Summary of the Drude components used to calculate the PAH cross sections for the model of T Cha.}
                        \label{tab:pahresonance}
                        \begin{tabular}{lcccc}
                                \hline
                                &                      &                     & \multicolumn{2}{c}{$\sigma_{\mathrm{abs}j}=\int\sigma_{\mathrm{abs},j}{d\lambda}^{-1}$}\\
                                \cline{4-5}
                                j  & $\lambda_\mathrm{j}$ & $\gamma_\mathrm{j}$ & Neutral           & Ionised \\
                                & ($\mu$m)             &                     & ($10^{-20}$ cm/C) & ($10^{-20}$ cm/C) \\
                                \hline 
                                1 & 0.0722 & 0.195 & $7.97\times10^7$ & $7.97\times10^7$\\ 
                                2 & 0.2175 & 0.217 & $1.23\times10^7$ & $1.23\times10^7$\\ 
                                3 & 1.05 & 0.055 & 0 & $2.0\times10^4$\\ 
                                4 & 1.26 & 0.11 & 0 & $7.8\times10^3$\\ 
                                5 & 1.905 & 0.09 & 0 & -146.5\\ 
                                6 & 3.3 & 0.012 & 394(H/C) & 89.4(H/C)\\ 
                                7 & 5.27 & 0.034 & 2.5 & 20.0\\ 
                                8 & 5.7 & 0.035 & 4.0 & 32.0\\
                                9$^*$ & 6.021 & 0.014 & 2.47 & 19.7\\ 
                                10$^*$ & 6.182 & 0.011 & 3.14 & 25.0\\ 
                                11$^*$ & 6.27 & 0.023 & 23.8 & 190\\ 
                                12 & 6.69 & 0.07 & 7.35 & 59.0\\ 
                                13$^*$ & 7.586 & 0.025 & 4.98 & 44.8\\ 
                                14$^*$ & 7.89 & 0.046 & 9.4 & 84.6\\ 
                                15$^*$ & 8.143 & 0.067 & 52.3 & 418\\ 
                                16$^*$ & 8.65 & 0.035 & 18.6(H/C) & 130(H/C)\\ 
                                17 & 10.68 & 0.02 & 0.3(H/C) & 0.3(H/C)\\ 
                                18$^*$ & 11.22 & 0.0075 & 14.3(H/C) & 13.5(H/C)\\ 
                                19$^*$ & 11.288 & 0.012 & 26.1(H/C) & 24.6(H/C)\\ 
                                20$^*$ & 11.428 & 0.025 & 30.5(H/C) & 28.8(H/C)\\ 
                                21 & 11.99 & 0.045 & 24.2(H/C) & 20.5(H/C)\\ 
                                22$^*$ & 12.62 & 0.042 & 21.5(H/C) & 20.5(H/C)\\ 
                                23 & 12.69 & 0.013 & 1.3(H/C) & 1.3(H/C)\\ 
                                24 & 13.48 & 0.04 & 8.0(H/C) & 8.0(H/C)\\ 
                                25 & 14.19 & 0.025 & 0.6 & 0.6\\ 
                                26 & 15.9 & 0.02 & 0.04 & 0.04\\ 
                                27 & 16.447 & 0.014 & 0.5 & 0.5\\ 
                                28 & 17.04 & 0.065 & 2.99 & 2.99\\ 
                                29 & 17.375 & 0.012 & 0.15 & 0.15\\ 
                                30 & 17.87 & 0.016 & 0.09 & 0.09\\ 
                                31 & 18.92 & 0.019 & 0.1 & 0.1\\ 
                                32 & 15.0 & 0.8 & 21.3 & 21.3\\  
                                \hline
                        \end{tabular}
                        \tablefoot{The parameters are listed in the same format as in Table 4 of \cite{Draine+2021}. The components specifically calculated for this work are marked with an asterisk.}
                \end{table}
                
                The astro-PAH opacities (grain cross sections) were empirically derived in order to fit the main C-C and C-H vibrational modes that give rise to the observed AIBs, including most common ones at 3.3, 6.2, 7.7, 8.6, and 11.3 $\mu$m \citep{LiDraine2001}. These opacities were also mainly based on the observations of the most generalised form of PAH spectral template that arise from the ISM, including reflection nebulae and galaxies, hence corresponded to class $\mathcal{A}$ profiles. 
                Spectral decomposition of a large sample of galaxy spectra with prominent AIBs by \cite{Smith+2007} suggested a few more sub-modes: the 7.7 $\mu$m band could be composed of three modes at 7.417, 7.598, and 7.850 $\mu$m; the 11.3 $\mu$m band can be fitted with two modes at 11.23 and 11.30 $\mu$m. The 6.2 and 8.6 $\mu$m bands were fitted with single modes centred at 6.22 and 8.61 $\mu$m, respectively. An additional modes between the 7.7 and 8.6 $\mu$m bands wad proposed to better fit the observations. All these information and the results from new observations in the post-Spitzer era were incorporated in modifying the empirical astro-PAH opacities \citep{DraineLi2007, Draine+2021}.  
                
                The class $\mathcal{A}$ template significantly differs from the class $\mathcal{C}$, particularly in the 7-9 $\mu$m region. Therefore, to replicate the class $\mathcal{C}$ signatures in the AIB profiles of T Cha, we modify the parameters of the relevant Drude components in Table 4 of \cite{Draine+2021}. We note that we only modulate the parameters that alter the shape of the generated profiles, and not the original astro-PAH model. 
                
                Based on \cite{LiDraine2001}, the absorption cross section per carbon atom contributed by the $j$th Drude component is     
                \begin{equation}
                        C^{\mathrm{PAH}}_{\mathrm{abs},j}(\lambda)/N_\mathrm{C} = \frac{2}{\pi}\frac{\gamma_j\lambda_j\sigma_{\mathrm{int},j}}{\left( \lambda/\lambda_j - \lambda_j/\lambda \right)^2 + \gamma_j^2},\\  
                        \label{eqn:B1}
                \end{equation}          
                where, $\sigma_{\mathrm{int},j}$ is the integrated absorption strength of the $j$th profile, $N_\mathrm{C}$ denotes the number of C atoms, and the rest of the symbols are the same as in Equation \ref{eqn:1}. We note that $\sigma_{\mathrm{int},j}$ can be tuned for a number of Drude profile components such that it can replicate a given PAH template. Further, the choice of the $\sigma_{\mathrm{int},j}$ may be guided by $\lambda_\mathrm{j}$, $b_\mathrm{j}$, and $\gamma_\mathrm{j}$ corresponding to each profile.
                Comparing Equation \ref{eqn:1} and Equation \ref{eqn:B1}, we derive $\sigma_{\mathrm{int},j}$ for all the Drude components calculated for the 6.2, 8.1 and 11.3 $\mu$m AIBs listed in Table \ref{tab:pahflux}, in a relative scale, using the values of $\lambda_\mathrm{j}$, $b_\mathrm{j}$, and $\gamma_\mathrm{j}$. 
                Then, we scale the total $\sigma_{\mathrm{int},j}$ values individually summed for the 6.2, 8.1 and 11.3 $\mu$m bands to the absolute $\sigma_{\mathrm{int},j}$ scales corresponding to the 6.2 $\mu$m band, the 7.7+8.6 complex, and the 11.3 $\mu$m band, respectively, using the values given by \cite{Draine+2021}{\footnote{For example, in case of the 8.1 $\mu$m band, the summed cross sections of the 7.586, 7.89, 8.143, and 8.65 $\mu$m components calculated in this work matches the summed cross sections of the 7.417, 7.598, 7.85, 8.33, and 8.61 $\mu$m components, corresponding to the 7.7+8.6 $\mu$m complex, given in \cite{Draine+2021}}}. The scaling is first applied to get the $\sigma_{\mathrm{int},j}$ values for the neutral PAHs in the required absolute units, and then the values were further scaled with appropriate factors to obtain the $\sigma_{\mathrm{int},j}$ values for ionised PAHs. Since some components carries the dependency on the hydrogen-to-carbon ratio ($\mathrm{H/C}$, see \citealt{DraineLi2007, Draine+2021}), during the scaling process, we adopt $\mathrm{H/C}=0.375$ (also see below). 
                
                Following this process, we obtain the cross sections for 30 different sizes of PAHs spanning the size range of $\sim3-50$ \AA, corresponding to $N_\mathrm{C}\sim10-50000$. The cross-sections for each PAH sizes are calculated in a wavelength range of 1 nanometre to 1 millimetre. Fig. \ref{fig:cabs} shows a comparison of the neutral and ionised PAH absorption cross section per C atom given by \cite{Draine+2021} and that calculated in this work corresponding to $N_\mathrm{C}=24$. We finally configure the files including the cross section data following the required format by {\tt MCFOST} in order for them to be read by the code during radiative transfer calculations. To run the calculations, it is also necessary to supply the size distribution, mass fraction of the dust species, and the total dust mass in a region, from which the code can internally calculate the opacities for individual species. Thus, the calculated PAH cross sections are are translated to opacities within the code to generate the PAH emission bands in the modelled spectrum that can be compared to the observed AIBs.  
                
                Table \ref{tab:pahresonance} enlist all the Drude components used to derive the PAH cross section similar to that given in Table 4 of \cite{Draine+2021}, where the components calculated for this work are marked with asterisk. The 6.2 {$\mu$m} band is represented with three components at 6.021, 6.182, and 6.27 $\mu$m, compared to a single component at 6.22 $\mu$m assigned earlier for galaxy spectra. The 8.1 $\mu$m band is represented with four components at 7.586, 7.89, 8.143, and 8.65 $\mu$m, combining 7.7 and 8.6 $\mu$m bands, to replicate the class $\mathcal{C}$ features. The 11.3 $\mu$m band has three components at 11.22, 11.288, and 11.428 $\mu$m. 
                
                The dependency of the hydrogen-to-carbon ratio of some components are adopted as similar to that in \cite{DraineLi2007}, where a factor $\mathrm{H/C}$ is {associated to the cross sections} of the components: $\mathrm{H/C}=0.5$, for $N_\mathrm{C}\leq25$; $\mathrm{H/C}=0.5/\sqrt{N_\mathrm{C}/25}$, for $25\leq{N_\mathrm{C}}\leq100$; and $\mathrm{H/C}=0.25$, for $N_\mathrm{C}\geq100$.  
                
                \FloatBarrier   
        
            \section{Peak-to-peak flux estimation of the AIBs} \label{app:peaktopeakflux}
            
            To estimate the peak-to-peak flux of each AIB profile, we first define two base points towards the blue and red wings of the profile, where the profile meets the continuum, and measure their fluxes. We join the two points with a straight line that approximates the continuum (base) level. We then use linear interpolation between the two base points and obtain the base flux corresponding to the peak wavelength of the profile. Hence, the difference between the peak absolute flux and the base flux gives the peak-to-peak flux of the profile. 
            
            \begin{figure}[h]
                \centering
                \includegraphics[width=\columnwidth]{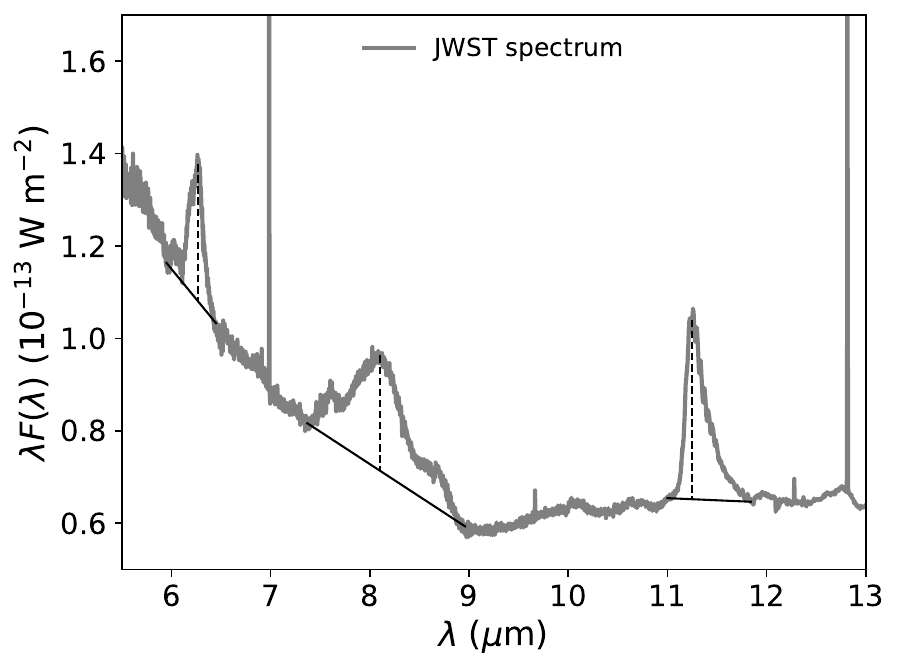}
                \caption{Graphical representation of the peak-to-peak flux estimation of the AIBs for the fitting procedure discussed in Sect. \ref{sec:pahabundancegrid}. A linear continuum baseline under each AIB is assumed, shown in black solid line. The black dashed lines connect the profile peaks with point at the baseline corresponding to the peak wavelength.}
                \label{fig:peaktopeakflux}      
            \end{figure}
        
        \FloatBarrier
                        
                \section{Supplementary model results} \label{app:modelset}      
                
\begin{figure}[h]
                \centering
                \includegraphics[width=\columnwidth]{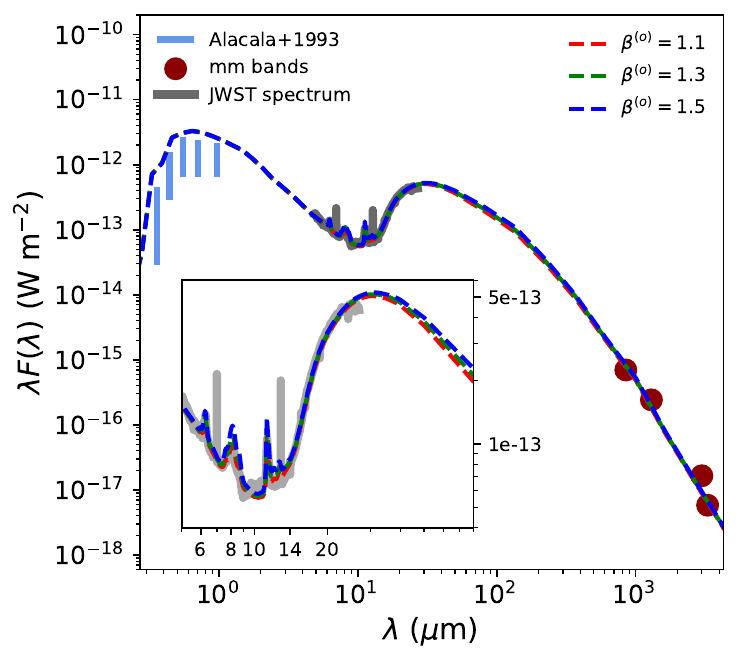}
                \caption{Variation in the modelled spectrum with the flaring index of the outer disk ($\beta^{(o)}$), with other parameters were kept fixed to their best fiducial model values presented in this paper. The full SED shows no considerable differences in the optical and millimetre fluxes with $\beta$. Variation due to $\beta^{(o)}$ mainly arises in the FIR wavelengths within $\sim30-500$ $\mu$m and in the modelled PAH fluxes (see Sect. \ref{sec:resultsdiscussion}). We note that the scale height is defined at 25 au, hence, the model with higher $\beta^{(o)}$ would have a larger scale height at the inner edge of the outer disk at 31 au. For an appropriate comparison among different models with varying $\beta^{(o)}$, so that they have the same scale heights at 31 au, we re-adjust the scale heights at 25 au. For the models with $\beta^{(o)}=1.3$ and $\beta^{(o)}=1.5$, the adjusted scale heights at 25 au will be 1.82 and 1.745 au, respectively).}
                \label{fig:modelsetbeta}        
        \end{figure}

\begin{figure}[h]
                \centering
                \includegraphics[width=\columnwidth]{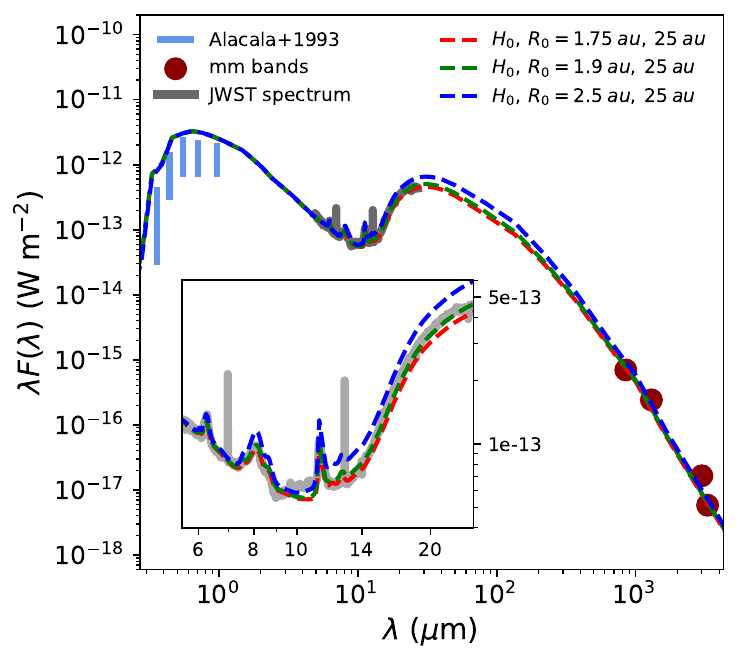}
                \caption{Variation in the modelled spectrum with the scale height ($H_0/R_0$) of the outer disk defined at 25 au, with other parameters were kept fixed to their best fiducial model values presented in this paper.}
                \label{fig:modelsetscaleheight} 
        \end{figure}

\begin{figure}[h]
                \centering
                \includegraphics[width=\columnwidth]{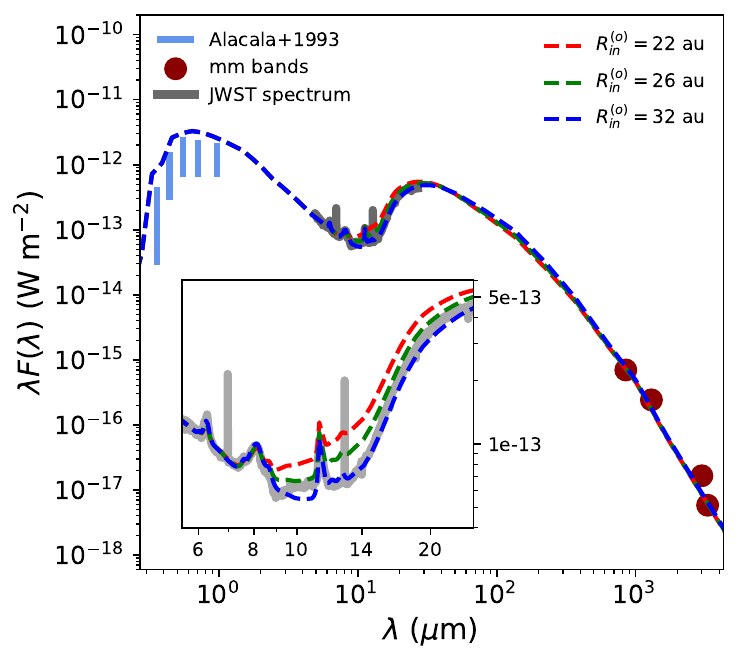}
                \caption{Variation in the modelled spectrum with the inner edge of the outer disk ($R^{(o)}_\mathrm{in}$), with other parameters were kept fixed to their best fiducial model values presented in this paper.}
                \label{fig:modelsetinnerradouter}       
            \end{figure}

\begin{figure}[h]
                \centering
                \includegraphics[width=\columnwidth]{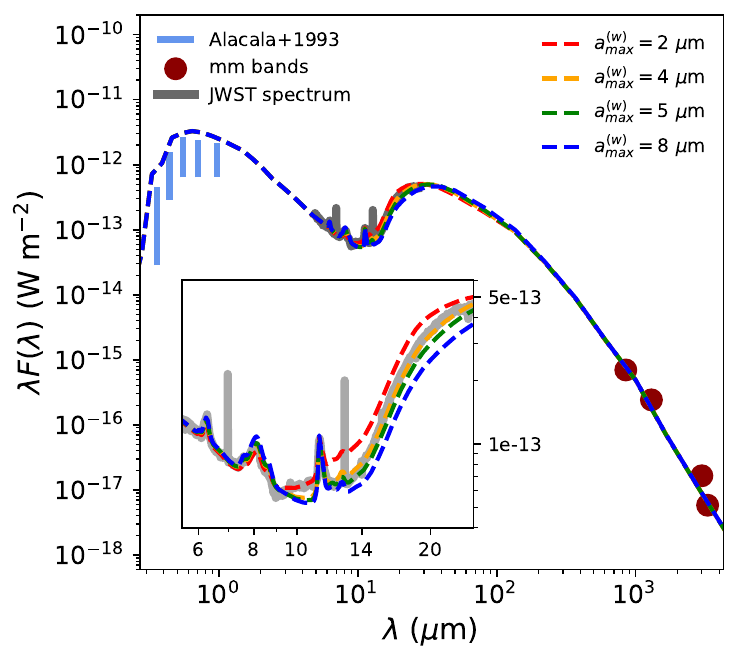}
                \caption{Variation in the modelled spectrum with the maximum grain size in the wall ($a^{(w)}_\mathrm{max}$), with other parameters were kept fixed to their best fiducial model values presented in this paper.}
                \label{fig:modelsetwallwidth}   
            \end{figure}

\begin{figure}[h]
                \centering
                \includegraphics[width=\columnwidth]{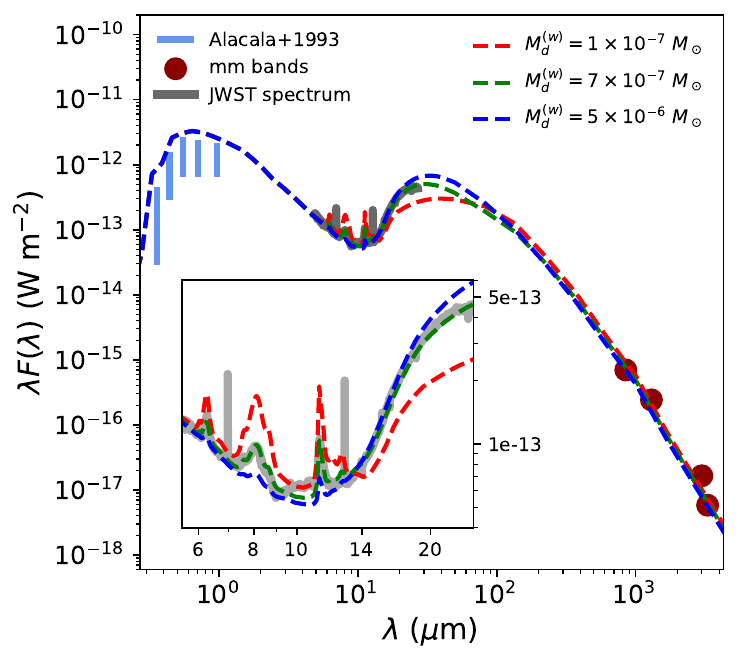}
                \caption{Variation in the modelled spectrum with the dust mass of the outer disk wall ($M^{(w)}_\mathrm{d}$), with other parameters were kept fixed to their best fiducial model values presented in this paper.}
                \label{fig:modelsetdustmasswall}        
            \end{figure}

\begin{figure}[h]
                \centering
                \includegraphics[width=\columnwidth]{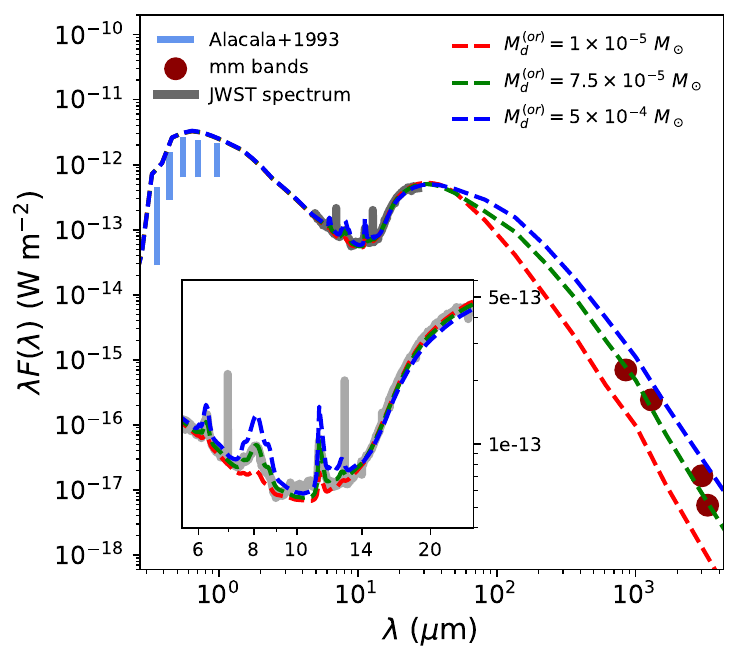}
                \caption{Variation in the modelled spectrum with the dust mass of the outer region of the outer disk ($M^{(or)}_\mathrm{d}$), with other parameters were kept fixed to their best fiducial model values presented in this paper.}
                \label{fig:modelsetdustmassouter}       
            \end{figure}

\begin{figure}[h]
                \centering
                \includegraphics[width=\columnwidth]{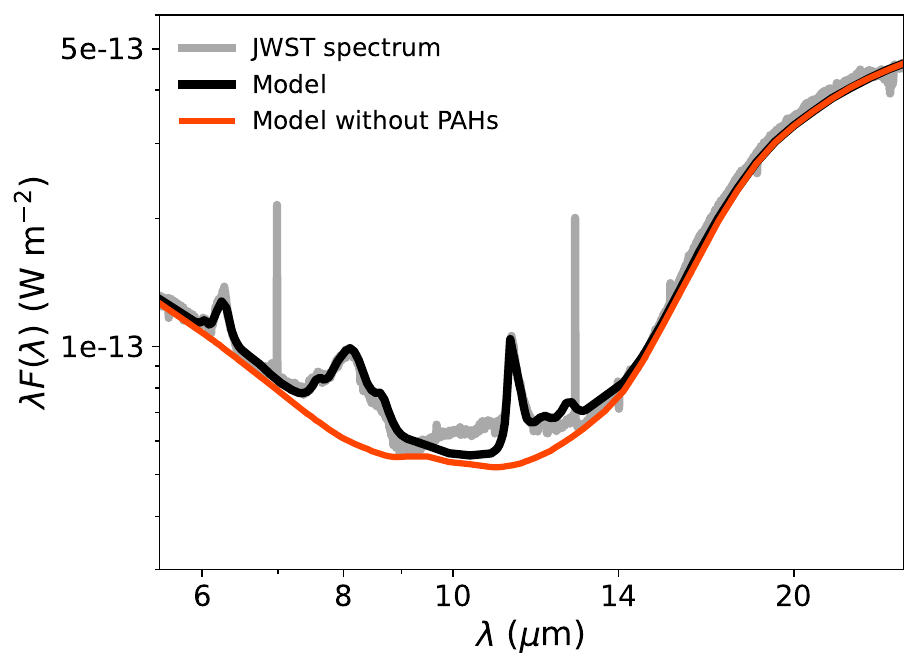}
                \caption{Comparison of the T Cha disk model computed in this paper (black) and the same model, but excluding the contribution of the PAHs to the modelled spectrum (red). Hence, this plot depicts the relative contribution of the PAHs to the thermal dust continuum.}
                \label{fig:pah}
        \end{figure}

\begin{figure}[h]
                \centering
                \includegraphics[width=\columnwidth]{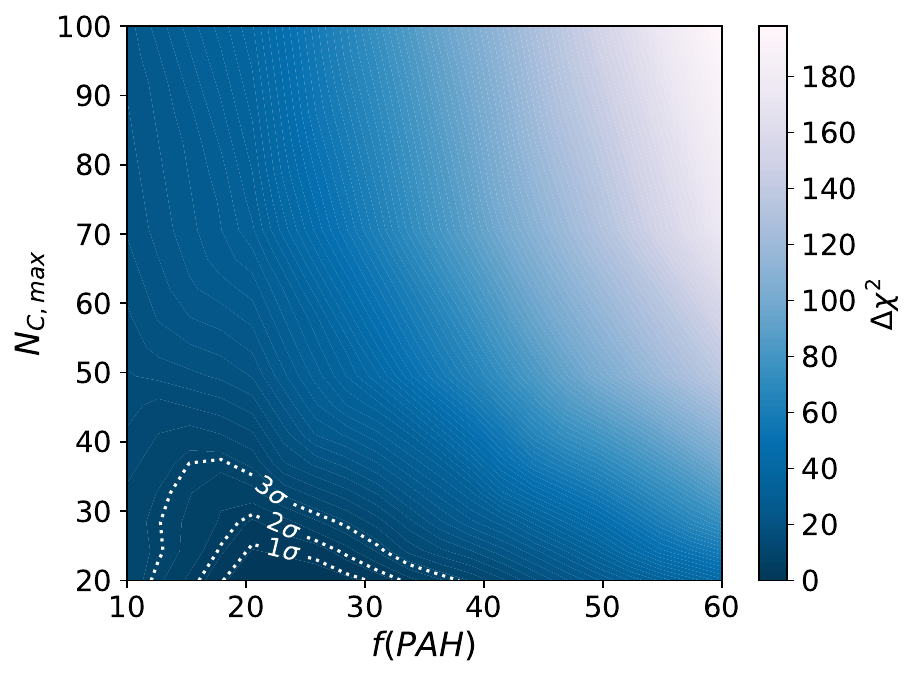}
                \caption{$\Delta\chi^2$ contours of the fits to the observed 6.2/8.1, 6.2/11.3, and 8.1/11.3 $\mu$m AIB ratios in $N_\mathrm{C, max}-f_\mathrm{PAH}$ parameter space (for $N_\mathrm{C, min}=10$ and $\phi=0.1$ corresponding to the minimum $\chi^2$ model, see text in Sect. \ref{sec:pahsizechargemass}).}
                \label{fig:pahsizemass}
        \end{figure}

\begin{figure}[h]
                \centering
                \includegraphics[width=\columnwidth]{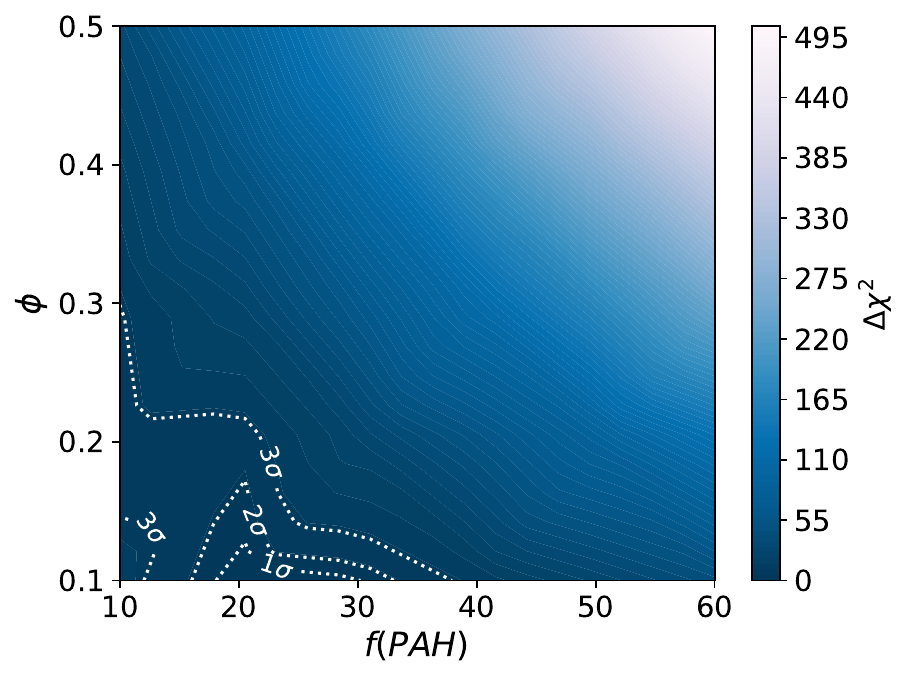}
                \caption{$\Delta\chi^2$ contours of the fits to the observed 6.2/8.1, 6.2/11.3, and 8.1/11.3 $\mu$m AIB ratios in $\phi-f_\mathrm{PAH}$ parameter space (for $N_\mathrm{C, min}=10$ and $N_\mathrm{C, max}=20$ corresponding to the minimum $\chi^2$ model, see text in Sect. \ref{sec:pahsizechargemass}).}
                \label{fig:pahchargemass}
        \end{figure}

                In this section, we provide model results that may serve as supplements to the main results presented in the paper. We show the dependency of the models on a particular parameter of the outer disk. Models are generated by varying one parameter, while the other best fiducial model parameters are kept fixed. The results are shown in Fig. \ref{fig:modelsetbeta} to Fig. \ref{fig:modelsetdustmassouter}. The specific parameter we varied could be found from the caption of each figure. We provide a plot in Fig. \ref{fig:pah} that depicts the contribution of the PAHs to the thermal continuum in our best fiducial model. Fig. \ref{fig:pahsizemass} and Fig. \ref{fig:pahchargemass} show the $\Delta\chi^2$ contours corresponding to the PAH size, ionisation fraction, and PAH-to-dust mass ratio parameter spaces explored in Sect. \ref{sec:pahsizechargemass}. 
                        
        \end{appendix}
        
\end{document}